\documentclass[10pt,journal,compsoc]{IEEEtran}

\ifCLASSOPTIONcompsoc

  \usepackage[nocompress]{cite}
\else
  \usepackage{cite}
\fi
\usepackage{longtable}
\usepackage{booktabs}
\usepackage{amsmath}
\usepackage{graphicx}
\usepackage{mwe}
\usepackage{graphbox}
\usepackage{amsfonts} 
\usepackage{enumitem}
\usepackage{amssymb}
\usepackage{ragged2e}
\usepackage[colorlinks, linkcolor=blue, anchorcolor=blue, citecolor=blue]{hyperref}
\usepackage[numbers,sort&compress]{natbib}
\usepackage{comment}
\usepackage{color}
\usepackage{bbding} 
\usepackage{diagbox} 
\usepackage{stfloats}
\usepackage{float}
\usepackage{multirow}
\ifCLASSINFOpdf

\else

\fi

\begin{document}

\title{QSAN: A Near-term Achievable Quantum Self-Attention Network}

\author{Jinjing Shi\textsuperscript{\#},~\IEEEmembership{Member,~IEEE,}
	Ren-Xin Zhao\textsuperscript{\#},~\IEEEmembership{Member,~IEEE,}
	Wenxuan Wang,~\IEEEmembership{Student Member,~IEEE,}
	Shichao Zhang\textsuperscript{*},~\IEEEmembership{Senior Member,~IEEE,}
    and~Xuelong~Li,~\IEEEmembership{Fellow,~IEEE}
    
\IEEEcompsocitemizethanks{
	
	\IEEEcompsocthanksitem{Jinjing Shi, Ren-Xin Zhao, Shichao Zhang and Wenxuan Wang are with the School of Computer Science and Engineering, Central South Univerisity, China, Changsha, 410083.}
\IEEEcompsocthanksitem{Xuelong Li is with the School of Artificial Intelligence, OPtics and ElectroNics (iOPEN), Northwestern Polytechnical University, Xi'an 710072, P. R. China. Xuelong Li is also with the Key Laboratory of Intelligent Interaction and Applications (Northwestern Polytechnical University), Ministry of Industry and Information Technology, Xi'an 710072, P. R. China.} 

\IEEEcompsocthanksitem{Jinjing Shi and Ren-Xin Zhao contributed equally to this work. Shichao Zhang is corresponding author.} 
\IEEEcompsocthanksitem{E-mails: shijinjing@csu.edu.cn, 13061508@alu.hdu.edu.cn, zhangsc@cs-
	u.edu.cn, wangwenxuan0524@csu.edu.cn, li@nwpu.edu.cn.} 
}
\thanks{Manuscript received XXXX; revised XXXX.}
}


\IEEEtitleabstractindextext{%
\begin{abstract}\justifying
Self-Attention Mechanism (SAM) is good at capturing the internal connections of features and greatly improves the performance of machine learning models, espeacially requiring efficient characterization and feature extraction of high-dimensional data. A novel Quantum Self-Attention Network (QSAN) is proposed for image classification tasks on near-term quantum devices. First, a Quantum Self-Attention Mechanism (QSAM) including Quantum Logic Similarity (QLS) and Quantum Bit Self-Attention Score Matrix (QBSASM) is explored as the theoretical basis of QSAN to enhance the data representation of SAM. QLS is employed to prevent measurements from obtaining inner products to allow QSAN to be fully implemented on quantum computers, and QBSASM as a result of the evolution of QSAN to produce a density matrix that effectively reflects the attention distribution of the output. Then, the framework for one-step realization and quantum circuits of QSAN are designed for fully considering the compression of the measurement times to acquire QBSASM in the intermediate process, in which a quantum coordinate prototype is introduced as well in the quantum circuit for describing the mathematical relation between the output and control bits to facilitate programming. Ultimately, the method comparision and binary classification experiments on MNIST with the pennylane platform demonstrate that QSAN converges about 1.7x and 2.3x faster than hardware-efficient ansatz and QAOA ansatz respevtively with similar parameter configurations and 100\% prediction accuracy, which indicates it has a better learning capability. \textcolor{red}{QSAN is quite suitable for fast and in-depth analysis of the primary and secondary relationships of image and other data, which has great potential for applications of quantum computer vision from the perspective of enhancing the information extraction ability of models.}
\end{abstract}

}

\maketitle

\IEEEdisplaynontitleabstractindextext

\IEEEpeerreviewmaketitle

\IEEEraisesectionheading{\section{Introduction}\label{sec:introduction}}

\IEEEPARstart{I}{N} recent years, tremendous progress has been achieved in the field of machine learning \cite{5.32}, where SAM is an important machine learning operator that produces attention scores from individual sequence itself to calculate sequence. It was originally introduced by a deep learning framework for machine translation called Transformer to overcome the problem of long-range dependencies in previous neural networks such as RNNs \cite{0.0}. Corresponding experimental results demonstrate that SAM can reduce the dependence on external information and better capture the intrinsic relevance of features \cite{0.0}. This importance has been repeatedly evidenced in many fields such as computer vision \cite{0.3,0.31,0.311,0.312}, natural language processing \cite{0.1,0.111,0.11}, speech \cite{0.2} and emotion analysis \cite{0.21}. For example, in 2021, 84.7\% first accuracy on the ImageNet benchmark was realized by a BoTNet with SAM \cite{0.313}. In the same year, this accuracy was boosted to 87.3\% by a Swin Transformer with shifted window SAM \cite{0.32}. It is a remarkable progress compared to the models before the advent of SAM, while it seems to be not efficient enough for high-dimensional data representation and feature extraction. Thus it leads us to think whether there is a new platform that can enhance capabilities of SAM in data characterization and how effective it is.

Fortunately, a feasible solution to this problem can be supplied by quantum computer. Quantum computer is considered as a new paradigm that can achieve quadratic speedup of algorithms, which has made significant breakthroughs in recent years \cite{0.33,0.34,1,2,3}. The superiority offered by quantum computers, also known as quantum supremacy \cite{4,5}, specifically refers to the exponential storage and secondary computational acceleration arising from the effects of quantum properties, as reflected in Quantum Machine Learning (QML) as well as quantum simulations \cite{5.01,5.02}. In particular, the talent of quantum computers in data representation \cite{5.03,30} led us to ponder whether it can efficiently express SAM for sequences and how to embody quantum advantages in SAM, which has already inspired some exploratory works. For instance, in 2017, 
Niu \textit{et al.} exploited the idea of weak measurement in quantum mechanics to construct a parameter-free, more efficient quantum attention \cite{5.0}, which is used in the LSTM framework and found to have better sentence modeling performance. In 2021, Zhao \textit{et al.} considered the quantum attention mechanism as a density matrix to construct more powerful sentence representations \cite{5.1}. Unfortunately, the above two approaches only involve certain physical concepts in quantum mechanics without providing specific quantum circuits. A recent meaningful effort was contributed by the Baidu group, where a Gaussian projection-based neural network using Variational Quantum Algorithm (VQA) \cite{5.3} to build ansatz \cite{5.31,5.32,5.321,5.33} on Noisy Intermediate-Scale Quantum (NISQ) \cite{5.2} devices was applied to text classification \cite{5.11}. In the current mainstream quantum-based SAM \cite{5.0,5.1,5.11}, only some parts of the SAM task are undertaken by the quantum computer, while it is more ideal for quantum computers to accomplish all the tasks of SAM, including calculating the attention scores and deriving outcomes in one step. Furthermore, the compression of the number of measurements is not sufficiently considered, where the more measurements are made, the more quantum data should be converted into classical data, resulting in more classical storage consumption.

A novel complete QSAN is proposed in order to enhance data characterization by exploring quantum advantages in SAM and more efficiently acquire  attention scores and outcomes in one step, where QSAM including QLS and QBSASM is explored as the theoretical basis of QSAN. Compared to SAM, QSAM demands exponentially less storage with the assistance of quantum representation for the same input sequence. In contrast to Refs. \cite{5.0,5.1,5.11}, QSAN can be potentially fully deployed and realized on quantum devices with fewer measurements. Moreover, the most encouraging thing is that we have exploringly discovered that young quantum computers may have quantum characteristic attention and can depict the distribution of outputs in a quantum language, but not to replace SAM or to beat all the schemes in the Refs. \cite{5.0,5.1,5.11}. Quantum characteristic here could be understood as probability, linearity and reversibility, while the quantum language refers to specialized terms in quantum mechanics, such as density matrix, Hamiltonian operators, etc. The main contributions of this paper are summarized as follows.

\begin{itemize} 
	\item QLS and QBSASM are defined to constitute the QSAM as the theoretical basis of QSAN to enhance the data representation, where QLS obtain the similarity directly by replacing the implicit inner product similarity with logical operations and QBSASM as a by-product generated from the evolution of QSAN can reflect the output attention distribution effectively in the form of a density matrix.

	\item The overall framework of QSAN and its quantum circuits are designed based on QSAM, in which quantum coordinates are proposed as an action criteria to simplify the design of QSAN to realize deriving the QBSASM and solution of task simultaneously in one step by compressing the number of measurements.
	
	\item Classification experiments for the MNIST are impletemented on the pennylane platform, which prove QSAN converges approximately 1.7x and 2.3x faster than hardware-efficient ansatz and QAOA ansatz respectively with 100\% prediction accuracy based on the same experimental setup and similar number of parameters, implying that QSAN has better learning capability.

\end{itemize} 

The rest of this paper is organized as follows. Basic theory and fundamental principles are summarized in section 2. QSAM with QLS, QBSASM are proposed in section 3. QSAN and its corresponding quantum circuits are designed in Section 4. The experiments and discussions are conducted in Section 5. Finally, the conclusion is drawn in Section 6.

\section{Preliminaries}

QML, SAM, VQA and quantum operators are briefly outlined in this section. 
\subsection{Quantum Machine Learning}
\textcolor{red}{QML is a new computational paradigm rooted in quantum mechanics.
	Various QML models with variational quantum algorithmic structures \cite{5.34}, such as quantum convolutional neural networks \cite{5.35}, quantum recurrent neural networks \cite{5.36} and quantum generative adversarial networks \cite{5.37}, have been introduced and applied in areas such as chemical analysis \cite{5.9} and physical dynamics simulation \cite{5.38}.
	QML has the ability to generate atypical data patterns via quantum mechanics to demonstrate potential quantum advantages \cite{5.39}, as verified by quantum kernel methods \cite{5.391,5.392}.
	In addition, some QML models exhibit quadratic acceleration with rigorous mathematical proofs \cite{5.393,5.394}.
	However, the development of QML is in the early stages, the situation of quantum resource constraints \cite{5.4} and quantum noise affecting accurate predictions \cite{5.41} make the performance of quantum advantages still elusive.
	Therefore, it is recommended to focus on enriching the use cases of QML to lay the foundation for the industrialization of future quantum computers, rather than pursuing quantum advantages alone \cite{5.42}.}

\subsection{Self-Attention Mechanism}
The input $\mathbf{In}=\{{{\mathbf{w}}_{i}}\}_{i=0}^{n-1}$, ${{\mathbf{w}}_{i}}\in {{\mathbb{R}}^{1\times l}}$ and the output $\mathbf{Out}=\{\mathbf{new\_}{{\mathbf{w}}_{j}}\}_{j=0}^{n-1}$, $\mathbf{new}\_{{\mathbf{w}}_{j}}\in {{\mathbb{R}}^{1\times l}}$ are defined, where $n$ is the total number of outputs, $l$ is the dimension of the element. $ {{\mathbf{w}}_{i}} $ ($ \mathbf{new}\_{{\mathbf{w}}_{j}} $) indicates the $ i $-th ($ j $-th) element of the sequence $ \mathbf{In} $ ($ \mathbf{Out}$). For conciseness, it is specified that all subscripts in the text are used to indicate the location of the variable in the sequence, except for special instructions.
Then SAM \cite{0.0} can be stated as
\begin{equation}\label{self_attent}
\mathbf{new}\_{{\mathbf{w}}_{i}}=\sum\limits_{j}{{{w}_{i,j}}{{\mathbf{V}}_{j}}}.
\end{equation}

In Eq. (\ref{self_attent}), $\mathbf{new}\_{{\mathbf{w}}_{i}}$ represents new output after the weighting operation. The weights
\begin{equation}\label{weight}
{{w}_{i,j}}=\text{softmax}\left( \frac{{{\mathbf{Q}}_{i}}\mathbf{K}_{j}^{\text{T}}}{\sqrt{d}} \right),
\end{equation}
also called attention scores, are obtained by normalizing the inner product ${{\mathbf{Q}}_{i}}\mathbf{K}_{j}^{\operatorname{T}}$. $\sqrt{d}$ is a scaling factor. \textcolor{red}{In Eq. (\ref{self_attent}) and Eq. (\ref{weight}),
	\begin{equation}\label{q1}
		{{\mathbf{Q}}_{i}}={{\mathbf{w}}_{i}}\cdot {{U}_{Q}},
	\end{equation} 
	\begin{equation}\label{v1}
		{{\mathbf{V}}_{j}}={{\mathbf{w}}_{j}}\cdot{{U}_{V}} ,
	\end{equation}
	and $\mathbf{K}_{j}^{\text{T}}$ are the query vector, the value vector and the transpose of the key vector
	\begin{equation}\label{k1}
		{{\mathbf{K}}_{j}}={{\mathbf{w}}_{j}}\cdot {{U}_{K}},
	\end{equation} where $\mathbf{w}_{i}$ and $\mathbf{w}_{j}$ are inputs. In Eq. (\ref{q1}) to Eq. (\ref{k1}), ${{U}_{Q}}$, ${{U}_{K}}$ and ${{U}_{V}}$ are three trainable parameter matrices named as query conversion matrix, key conversion matrix and value conversion matrix respectively. }

\subsection{Variational Quantum Algorithm}\label{vqa}

In the NISQ era, it is very difficult to fully deploy deep networks for deep learning on quantum computers with limited qubits. Firstly, the dimensionality of the model grows exponentially as the size of the quantum circuit gets larger \cite{5.4}. Secondly, noise imposes many unknowns on the
\begin{figure}[htbp]
	\vspace{0em}\centering
	\includegraphics[scale=0.23]{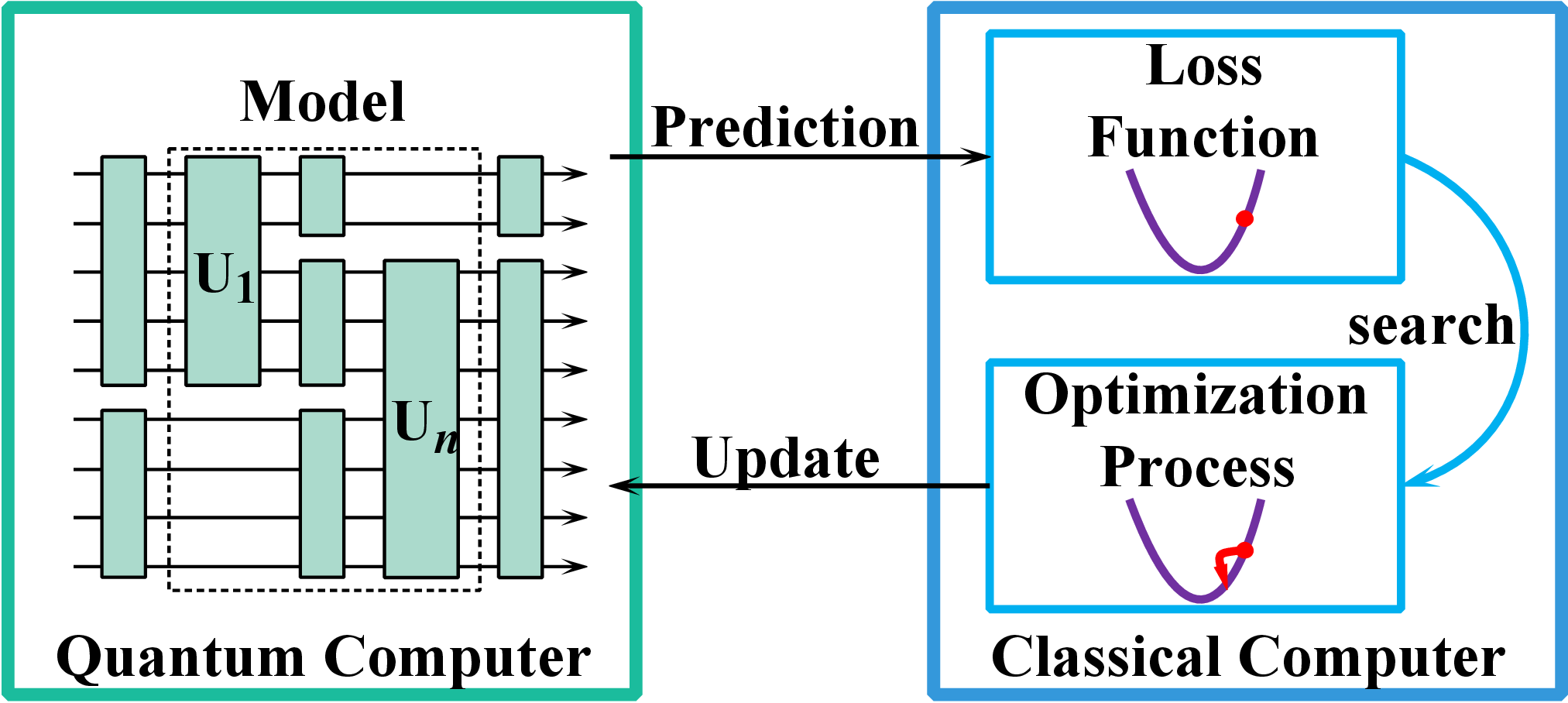}
	\vspace{0em}\caption{Framework of VQA}\label{VQA}
\end{figure} training results \cite{5.41,5.5}. Therefore, quantum-classical hybrid model is deemed as an efficient path. VQA is one such class of algorithms. The framework of VQA is exhibited in Fig. \ref{VQA}, which can be divided into two parts.

\textcolor{red}{The blue box designates the range of the classical computer.} This stage focuses on the calculation of the loss function and the optimization of the parameters, as shown in the two purple curves in Fig. \ref{VQA}. The general formulation of the loss function is
\begin{equation}\label{costs}
	\mathcal{C}(\theta )\text{=}\sum\limits_{k}{{{\mathcal{F}}_{k}}(\text{Tr}[{{\mathcal{O}}_{k}}\mathcal{U}{{(\theta )}_{{{\rho }_{k}}}}{{\mathcal{U}}^{\dagger }}(\theta )])},
\end{equation} 
\textcolor{red}{where $\mathcal{F}_{k}$ is a set of certain functions determined by specific tasks. Tr$[\cdot]$ indicates the trace of ${{\mathcal{O}}_{k}}\mathcal{U}{{(\theta )}_{{{\rho }_{k}}}}{{\mathcal{U}}^{\dagger }}(\theta )$. $\mathcal{O}_{k}$ is a set of observables. $\mathcal{U}{(\theta )}=\mathop{\otimes }_{i}{{U}_{i}}({{\theta }_{i}})$ denotes the product of a series of unitary operators, where $\mathbf{\theta }$ comprises a series of continuous or discrete hyperparameters. ${\mathcal{U}}^{\dagger }(\theta )$ is the conjugate transpose matrix of $\mathcal{U}{(\theta )}$. $\{{{\rho }_{k}}\}$ is the input state of the training set.}
 Currently, methods for optimizing Eq. (\ref{costs}) include iCANS \cite{5.6}, stochastic gradient descent \cite{5.7}, quantum natural gradient \cite{5.8}, etc.

\textcolor{red}{The green box stands for the quantum computer domain.} In this box, a ansatz model is drawn. The black dashed box is the centerpiece of this model, the ansatz, which is a circuit with a specific structure and function. Common examples of ansatz contain hardware-efficient ansatz \cite{5.9,5.10}, QAOA ansatz \cite{5.12,5.13, 5.14}, etc.

The arrows in Fig. \ref{VQA} illustrate the interaction of information of quantum computers and classical counterparts. The quantum computer provides the classical computer with quantum circuit measurements and loss function forms to be used for prediction. After the classical computer is trained, a new round of hyperparameters is uploaded and updated into the quantum circuit.

\subsection{Qubit and Operators}
In a quantum computer, the smallest element of information is a qubit $|\psi \rangle $ which is usually expressed as a linear superposition of two eigenstates $|0\rangle $ and $|1\rangle $ \cite{5.15}, namely 
\begin{equation}
	|\psi \rangle =\alpha |0\rangle +\beta |1\rangle,
\end{equation}
where $\alpha$ and $\beta$ as probability amplitudes, satisfy 
\begin{equation}
	|\alpha {{|}^{2}}+|\beta {{|}^{2}}=1.
\end{equation}
These qubits evolve through unitary operators $U$ which are also called quantum gates and refer to matrices that satisfy 
\begin{equation}
	\begin{matrix}
		{{U}^{-1}}={{U}^{\dagger }}, \\ 
		U{{U}^{\dagger }}=\mathcal{I}, \\ 
	\end{matrix}
\end{equation}
where $\mathcal{I}$ is the identity matrix, ${U}^{\dagger }$ is the complex conjugate of $U$. In this paper, rotating Pauli Y gate, Hadamard gate, SWAP gate, CNOT gate, Toffoli gate and multi-controlled Toffoli gate are mainly used, as shown in Tab. A1.

\section{Quantum Self-Attention Mechanism}\label{sec2}
This section presents a QSAM framework, in which QLS is proposed to measure logical similarity and enables QSAM to be freed from numerical operations such as addition, thus conserving more qubits. More importantly, QLS replaces the inner product similarity that needs to be implemented by measurement, which ensures that the task is always executed on the quantum computer without interruption. QBSASM derived from QLS expresses the weight distribution of the quantum computer on outputs in the form of a density matrix.

Firstly $\mathbf{In}$ and $\mathbf{Out}$ are re-expressed in quantum states as 
${{\mathbf{Q}}_{\text{in}}}=\{|{{\mathbf{w}}_{i}}\rangle \}_{i=0}^{n-1}$, $|{{\mathbf{w}}_{i}}\rangle \in {{\mathbb{R}}^{{{2}^{m}}\times 1}}$
and 
${{\mathbf{Q}}_{\text{out}}}=\{|\mathbf{new\_}{{\mathbf{w}}_{j}}\rangle \}_{j=0}^{n-1}$, $|\mathbf{new\_}{{\mathbf{w}}_{j}}\rangle \in {{\mathbb{R}}^{{{2}^{m+1}}\times 1}}$ 
respectively, where $m=\left\lceil {{\log }_{2}}l \right\rceil $ denotes both the number of qubits and the length of the quantum string. $l$ is the feature dimension of the classical output. Then QSAM is described as
\begin{equation}\label{QSA}
	|\mathbf{new}\_{{\mathbf{w}}_{i}}\rangle :=\underset{j}{\mathop{\odot }}\,\langle {{\mathbf{Q}}_{i}}|{{\mathbf{K}}_{j}}\rangle \otimes |{{\mathbf{V}}_{j}}\rangle .
\end{equation} 
where $\langle {{\mathbf{Q}}_{i}}|{{\mathbf{K}}_{j}}\rangle$ is QLS in subsection \ref{subQLS}. The symbol $\otimes $ signifies a tensor operation. The symbol $\odot $ encompasses two operations. One is to apply a multi-controlled Toffoli gate to several specific QLS elements. The method of selecting these specific QLS is called slicing operation in subsection \ref{QBSASM}. The other is to use CNOT gates to $|{{\mathbf{V}}_{j}}\rangle$ to perform dimensional compression. In Eq. (\ref{QSA}),
\begin{equation}\label{Qi}
	|{{\mathbf{Q}}_{i}}\rangle ={{U}_{q}}|{{\mathbf{W}}_{i}}\rangle,
\end{equation}
\begin{equation}\label{Kj}
	|{{\mathbf{K}}_{j}}\rangle ={{U}_{k}}|{{\mathbf{W}}_{j}}\rangle,
\end{equation}
and
\begin{equation}\label{Vj}
	|{{\mathbf{V}}_{j}}\rangle ={{U}_{v}}|{{\mathbf{W}}_{j}}\rangle,
\end{equation}
where $|{{\mathbf{Q}}_{i}}\rangle$, $|{{\mathbf{K}}_{j}}\rangle$ and $|{{\mathbf{V}}_{j}}\rangle$ are the elements of the query quantum state $|\mathbf{Q}\rangle$, the key quantum state $|\mathbf{K}\rangle$ and the value quantum state $|\mathbf{V}\rangle$, respectively. They can all be written in the form of quantum strings 
\begin{equation}
	|{{\mathbf{Q}}_{i}}\rangle =\underset{a=0}{\overset{m-1}{\mathop{\otimes }}}\,|{{\mathcal{Q}}_{i,a}}\rangle, 
\end{equation}
\begin{equation}
	|{{\mathbf{K}}_{j}}\rangle =\underset{b=0}{\overset{m-1}{\mathop{\otimes }}}\,|{{\mathcal{K}}_{j,b}}\rangle ,
\end{equation}
and 
\begin{equation}\label{Vjstring}
	|{{\mathbf{V}}_{j}}\rangle =\underset{c=0}{\overset{m-1}{\mathop{\otimes }}}\,|{{\mathcal{V}}_{j,c}}\rangle ,
\end{equation}
where $|{{\mathcal{Q}}_{i,a}}\rangle$, $|{{\mathcal{K}}_{j,b}}\rangle$ and $|{{\mathcal{V}}_{j,c}}\rangle$ indicate quantum superposition states at different positions. These representations are indicated in Fig. \ref{S5} and Fig. \ref{S8}. In Eq. (\ref{Qi}), Eq. (\ref{Kj}) and Eq. (\ref{Vj}), ${{U}_{q}}$, ${{U}_{k}}$ and ${{U}_{v}}$ are specified as three composite unitary operators with the identical structure but distinct parameters. The same composition means that all three matrices above are composed of $(m-1)$ Hadamard gates, $m$ rotating Pauli Y gates, and $m$ CNOT gates, and are arranged in order
\begin{equation}\label{UM}
	{{U}_{M\in \{q,k,v\}}}=CNO{{T}^{\otimes (m-1)}}{{R}_{y}}{{({{\theta }_{M}})}^{\otimes m}}{{H}^{\otimes m}}.
\end{equation}
The benefit of this design is to maintain that the probability amplitudes are all real numbers \cite{28}.  Furthermore, $|{{\mathbf{w}}_{j}}\rangle $ (or $|{{\mathbf{w}}_{i}}\rangle $) is the input quantum state that can be split into the form of the quantum string 
\begin{equation}
	|{{\mathbf{W}}_{j}}\rangle =\underset{i=0}{\overset{m-1}{\mathop{\otimes }}}\,|{{\mathcal{W}}_{j,i}}\rangle
\end{equation}
and this denotation is presented in Fig. \ref{S1}, Fig. \ref{S2} and Fig. \ref{S6}. 

Formally, Eq. (\ref{QSA}) is very similar to Eq. (\ref{self_attent}), but there are essential changes. Comparing Eq. (\ref{self_attent}) and Eq. (\ref{QSA}), Eq. (\ref{self_attent}) is an attention mechanism with nonlinear operations, while Eq. (\ref{QSA}) has a linearized, logical character, which makes it easier to be implemented on quantum computers across the board.
Futhermore, in Eq. (\ref{self_attent}), a large number of numerical operations are required, such as solving the inner product as well as weighted summation, which requires a large number of qubits to implement a network of numerical operations on existing quantum computers.
In contrast, Eq. (\ref{QSA}) reduces the implementation cost with QLS and saves even more qubits. \textcolor{red}{Finally, the relationship between QLS and $|{{\mathbf{V}}_{j}}\rangle $ in Eq. (\ref{QSA}) is a tensor product $\otimes $, not a product in Eq. (\ref{self_attent}), hence $|{{\mathbf{w}}_{i}}\rangle$ and $|\mathbf{new\_}{{\mathbf{w}}_{j}}\rangle$ differ in dimension.}

\subsection{Quantum Logical Similarity}\label{subQLS}

Inspired by CRC checksum and Frobenius inner product, QLS in the Definition 1 is proposed in order to obtain the similarity directly as well as to make QSAN avoid measurements in the intermediate process to ensure that quantum data are maintained during processing, which is quite different from the common way with SWAP test \cite{24} or Hadamard test \cite{25} to characterize the similarity between two quantum states $|{{\mathbf{Q}}_{i}}\rangle $ and $|{{\mathbf{K}}_{j}}\rangle $. 

\textbf{Definition 1} (QLS): For any quantum state $|\mathbf{Q}_{i}\rangle$ and $|\mathbf{K}_{j}\rangle$ with $i,j\in \{0,\cdots ,n-1\}$, QLS is redefined as
\begin{equation}\label{QK}
	\langle {{\mathbf{Q}}_{i}}|{{\mathbf{K}}_{j}}\rangle :=\underset{h}{\mathop{\oplus }}\,({{\mathcal{Q}}_{i,h}}\wedge {{\mathcal{K}}_{j,h}}),
\end{equation}
where $|{{\mathcal{Q}}_{i,h}}\rangle$ and $|{{\mathcal{K}}_{j,h}}\rangle$, $h\in \{0,\cdots ,m-1\}$ stand for the $h$-th qubit of $|\mathbf{Q}_{i}\rangle$ and $|\mathbf{K}_{j}\rangle$, respectively. The symbol $\oplus $ indicates modulo-two addition and the symbol $\wedge $ is logical AND operation. 
 
From the implementation point of view, AND operation and modulo-two addition can be realized with Toffoli gates and CNOT gates, respectively. Then Eq. (\ref{QK}) can be explained in terms of quantum gates:
\begin{equation}
	\begin{aligned}
		& Toffoli|{{\mathcal{Q}}_{i,h}},{{\mathcal{K}}_{j,h}},0\rangle \\ 
		& =|{{\mathcal{Q}}_{i,h}},{{\mathcal{K}}_{j,h}},{{\mathcal{Q}}_{i,h}}\wedge {{\mathcal{K}}_{j,h}}\rangle,  \\ 
		& CNOT|{{\mathcal{Q}}_{i,h}}\wedge {{\mathcal{K}}_{j,h}},{{\mathcal{Q}}_{i,h}}\wedge {{\mathcal{K}}_{j,h}}\rangle  \\ 
		& =|{{\mathcal{Q}}_{i,h}}\wedge {{\mathcal{K}}_{j,h}},({{\mathcal{Q}}_{i,h}}\wedge {{\mathcal{K}}_{j,h}})\oplus ({{\mathcal{Q}}_{i,h}}\wedge {{\mathcal{K}}_{j,h}})\rangle.  \\ 
	\end{aligned}
\end{equation}

After quantum logic calculations, the two lengthy qubit strings $|\mathbf{Q}_{i}\rangle$ and $|\mathbf{K}_{j}\rangle$ are compressed into a superposition state with QLS in Eq. (\ref{QK}), which forms the basic unit in the next subsection QBSASM.

\subsection{Quantum Bit Self-Attention Score Matrix}\label{QBSASM}
QBSASM, mathematically a matrix with QLS as elements, is a byproduct of the QSAN execution process to mirror the change in attention distribution before and after training.

It is known that the solution procedure for a single new output only is given by Eq. (\ref{QSA}). By analogy, the solution procedure for all outputs is depicted by the matrix form as follows:
\begin{equation}\label{rule}
\begin{aligned}
	& \left[ \begin{matrix}
		(\langle {{\mathbf{Q}}_{0}}|{{\mathbf{K}}_{0}}\rangle \otimes |{{\mathbf{V}}_{0}}\rangle )\odot \cdots \odot (\langle {{\mathbf{Q}}_{0}}|{{\mathbf{K}}_{n-1}}\rangle \otimes |{{\mathbf{V}}_{n-1}}\rangle )  \\
		(\langle {{\mathbf{Q}}_{1}}|{{\mathbf{K}}_{0}}\rangle \otimes |{{\mathbf{V}}_{0}}\rangle )\odot \cdots \odot (\langle {{\mathbf{Q}}_{1}}|{{\mathbf{K}}_{n-1}}\rangle \otimes |{{\mathbf{V}}_{n-1}}\rangle )  \\
		\vdots   \\
		(\langle {{\mathbf{Q}}_{n-1}}|{{\mathbf{K}}_{0}}\rangle \otimes |{{\mathbf{V}}_{0}}\rangle )\odot \cdots \odot (\langle {{\mathbf{Q}}_{n-1}}|{{\mathbf{K}}_{n-1}}\rangle \otimes |{{\mathbf{V}}_{n-1}}\rangle )  \\
	\end{matrix} \right] \\ 
	& =\left[ \begin{matrix}
		\text{ }|\text{ }\mathbf{new}\_{{\mathbf{w}}_{0}}\rangle   \\
		\text{ }|\text{ }\mathbf{new}\_{{\mathbf{w}}_{1}}\rangle   \\
		\vdots   \\
		\text{ }|\text{ }\mathbf{new}\_{{\mathbf{w}}_{n-1}}\rangle   \\
	\end{matrix} \right]. \\ 
\end{aligned}
\end{equation}
This led to the following definition of QBSASM.

\textbf{Definition 2} (QBSASM): The weight coefficient matrix
\begin{equation}\label{Matrix}
	\left[ \begin{matrix}
		\langle {{\mathbf{Q}}_{0}}|{{\mathbf{K}}_{0}}\rangle  & \langle {{\mathbf{Q}}_{0}}|{{\mathbf{K}}_{1}}\rangle  & \cdots  & \langle {{\mathbf{Q}}_{0}}|{{\mathbf{K}}_{n-1}}\rangle   \\
		\langle {{\mathbf{Q}}_{1}}|{{\mathbf{K}}_{0}}\rangle  & \langle {{\mathbf{Q}}_{1}}|{{\mathbf{K}}_{1}}\rangle  & \cdots  & \langle {{\mathbf{Q}}_{1}}|{{\mathbf{K}}_{n-1}}\rangle   \\
		\vdots  & {} & {} & \vdots   \\
		\langle {{\mathbf{Q}}_{n-1}}|{{\mathbf{K}}_{0}}\rangle  & \cdots  & \cdots  & \langle {{\mathbf{Q}}_{n-1}}|{{\mathbf{K}}_{n-1}}\rangle   \\
	\end{matrix} \right],
\end{equation}
is extracted from Eq. (\ref{rule}) as QBSASM to depict the distribution of attention scores, where each element is computed by QLS. 
The slicing operation mentioned previously comes into play here. Specifically, the slicing operation takes the elements QLS in each row of Eq. (\ref{Matrix}) (e.g., element $\langle {{\mathbf{Q}}_{0}}|{{\mathbf{K}}_{0}}\rangle $ to element $\langle {{\mathbf{Q}}_{0}}|{{\mathbf{K}}_{n-1}}\rangle $) as control bits of the multi-controlled quantum gate. The results of their operations are employed as new weights, thus reflecting the weighting operations of QSAN. In practice, QBSASM can be obtained by pennylane intercepting the density matrix of QLS.

In summary, in Eq. (\ref{Matrix}), since each element QLS from Eq. (\ref{QK}) is a superposition state rather than a scalar, QBSASM has higher dimensionality than the classical attention score matrix and can represent more state information. Particularly, as the dimensionality increases, QBSASM is more difficult to be  performed classical simulation, which is manifesting the storage advantage of quantum computers. 

\section{Quantum Self-Attention Network}

In this section, the overall framework of QSAN based on QSAM theory in the previous section and the corresponding quantum circuits are illustrated. Especially, a prototype of quantum coordinates is presented, which is a design guideline for quantum circuits with regular layout. The functional link between control bits and output bits can be established with the guidance of quantum coordinates to facilitate programming. It is also worth exploring in quantum circuit optimization.

\subsection{Framework of Quantum Self-Attention Network}
\begin{figure*}[htpb]
	\vspace{0em}\centering
	\includegraphics[scale=0.53]{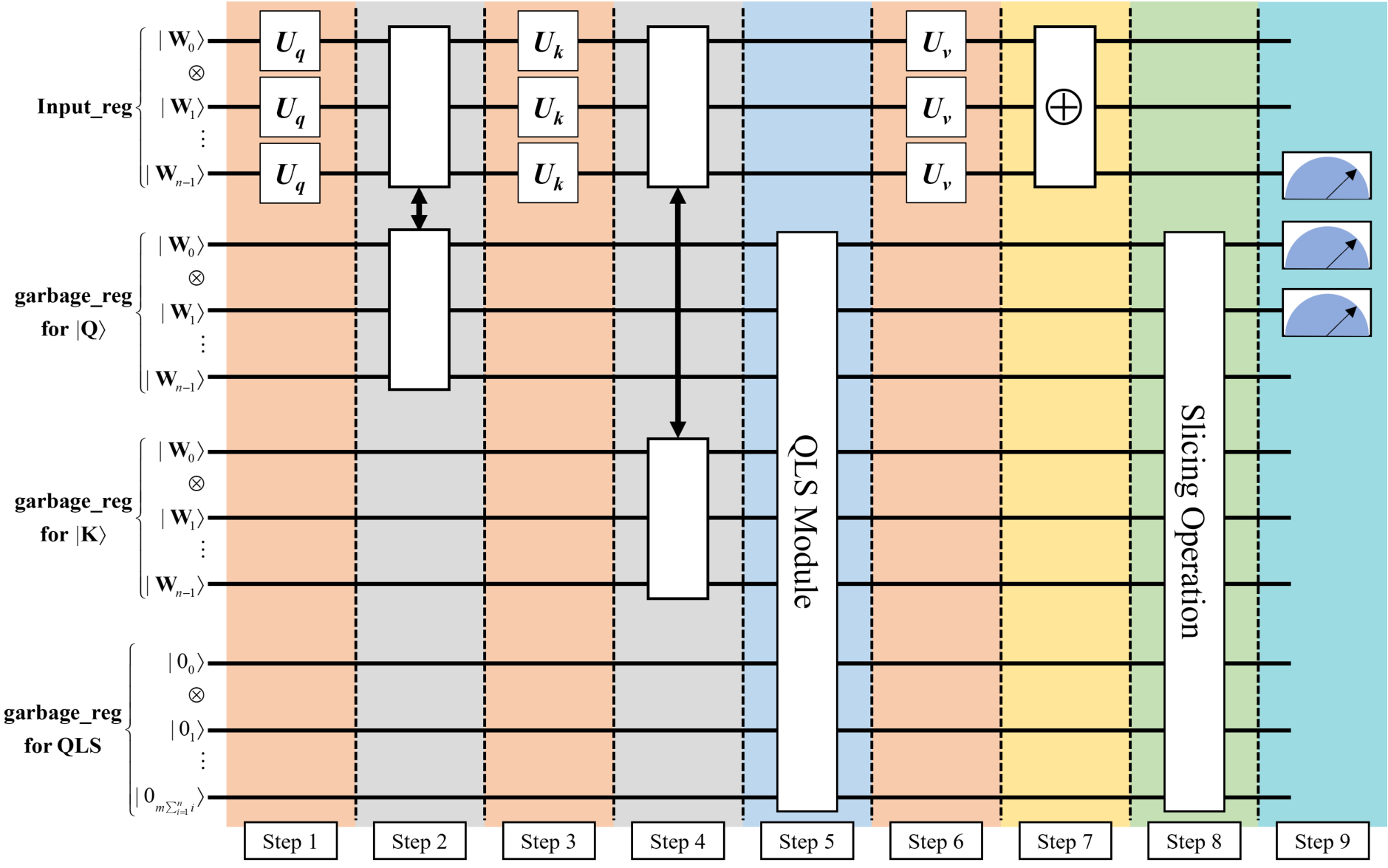}
	\vspace{0em}\caption{Circuit Model of QSAN to implement the QSAM theory in Eq. (\ref{rule}). Step 1, 3 and 6 are dedicated to calculate the query quantum state $|\mathbf{Q}\rangle $, the key quantum state $|\mathbf{K}\rangle $ and the value quantum state $|\mathbf{V}\rangle $, respectively. Steps 2 and 4 are barbell operations designed to swap with the corresponding garbage registers. Step 5 is the QLS module to compute the QLS elements, which produces the by-product QBSASM. Step 7 is the entanglement compression operation, which reduces the measurements. Step 8 is the slicing operation for calculating the final weights. The final step is measurement.}\label{QSAN}
\end{figure*}

To implement the QSAM theory in Eq. (\ref{rule}), the general framework of QSAN is designed in the style of Fig. \ref{QSAN}.

Vertically, QSAN in Fig. \ref{QSAN} consists of one input register and three garbage registers, where the three garbage registers are used to compute the query quantum state $|\mathbf{Q}\rangle$, the key quantum state $|\mathbf{K}\rangle$ and QLS, respectively. Moreover, the input register corresponds to \textbf{Input\_reg} in Fig. \ref{QSAN}, Fig. \ref{S1}, Fig. \ref{S2} and Fig. \ref{S6}, and the three garbage registers contrast to \textbf{garbage\_reg for $|\mathbf{Q}\rangle $}, \textbf{garbage\_reg for $|\mathbf{K}\rangle $} and \textbf{garbage\_reg for QLS} in Fig. \ref{QSAN}, Fig. \ref{S2}, Fig. \ref{S5} and Fig. \ref{S8}, respectively. In terms of resource consumption for QSAN in Fig. \ref{QSAN}, the first, second and third register takes $n\times m$ qubits each, while the fourth register needs $m\sum\nolimits_{i=1}^{n}{i}$ qubits, for a total of $3m\times n+m\sum\nolimits_{i=1}^{n}{i}$ qubits.

Horizontally, QSAN in Fig. \ref{QSAN} is divided into 9 steps in terms of workflow, including Step 1 to 8 and the final measurements. Functionally it is split into 5 special modules, mainly comprising barbell operation, QLS module and entanglement compression operation. In particular, the 5 special modules are marked with 5 different colors in Fig. \ref{QSAN}, where the same color means the same operation. Specifically, in Fig. \ref{QSAN}, Step 1, 3 and 6 are dedicated to calculate the query quantum state $|\mathbf{Q}\rangle $, the key quantum state $|\mathbf{K}\rangle $ and the value quantum state $|\mathbf{V}\rangle $, respectively. Steps 2 and 4 are barbell operations designed to swap with the corresponding garbage registers. Step 5 is the QLS module to compute the QLS elements, which produces the by-product QBSASM. Step 7 is the entanglement compression operation, which reduces the measurements. Step 8 is the slicing operation for calculating the final weights. The final step is measurement.
 
Additionally, for the successful execution of Eq. (\ref{rule}), for the first three registers in Fig. \ref{QSAN} (i.e., \textbf{Input\_reg}, \textbf{garbage\_reg for $|\mathbf{Q}\rangle $} and \textbf{garbage\_reg for $|\mathbf{K}\rangle $}), the same quantum coding \cite{30} is used to prepare the same initial quantum states 
\begin{equation}\label{inputs}
	|\mathbf{In}\rangle =\underset{j=0}{\overset{n}{\mathop{\otimes }}}\,|{{\mathbf{w}}_{j}}\rangle. 
\end{equation}

\subsection{Quantum Circuit}

In this subsection, the quantum circuit of QSAN is designed in detail following the steps in Fig. \ref{QSAN}.

\subsubsection{Quantum Coordinates}

In the design of quantum circuit for QSAN, a quantum coordinate is proposed for guiding the specific steps to implement a quantum circuit.

\textbf{{Definition 3}} (Quantum Coordinates): For a regularly arranged quantum circuit, the intersection of the number of layers and the circuit line number is the quantum coordinate.

\begin{figure}[htpb]
	\vspace{0em}\centering
	\includegraphics[scale=0.24]{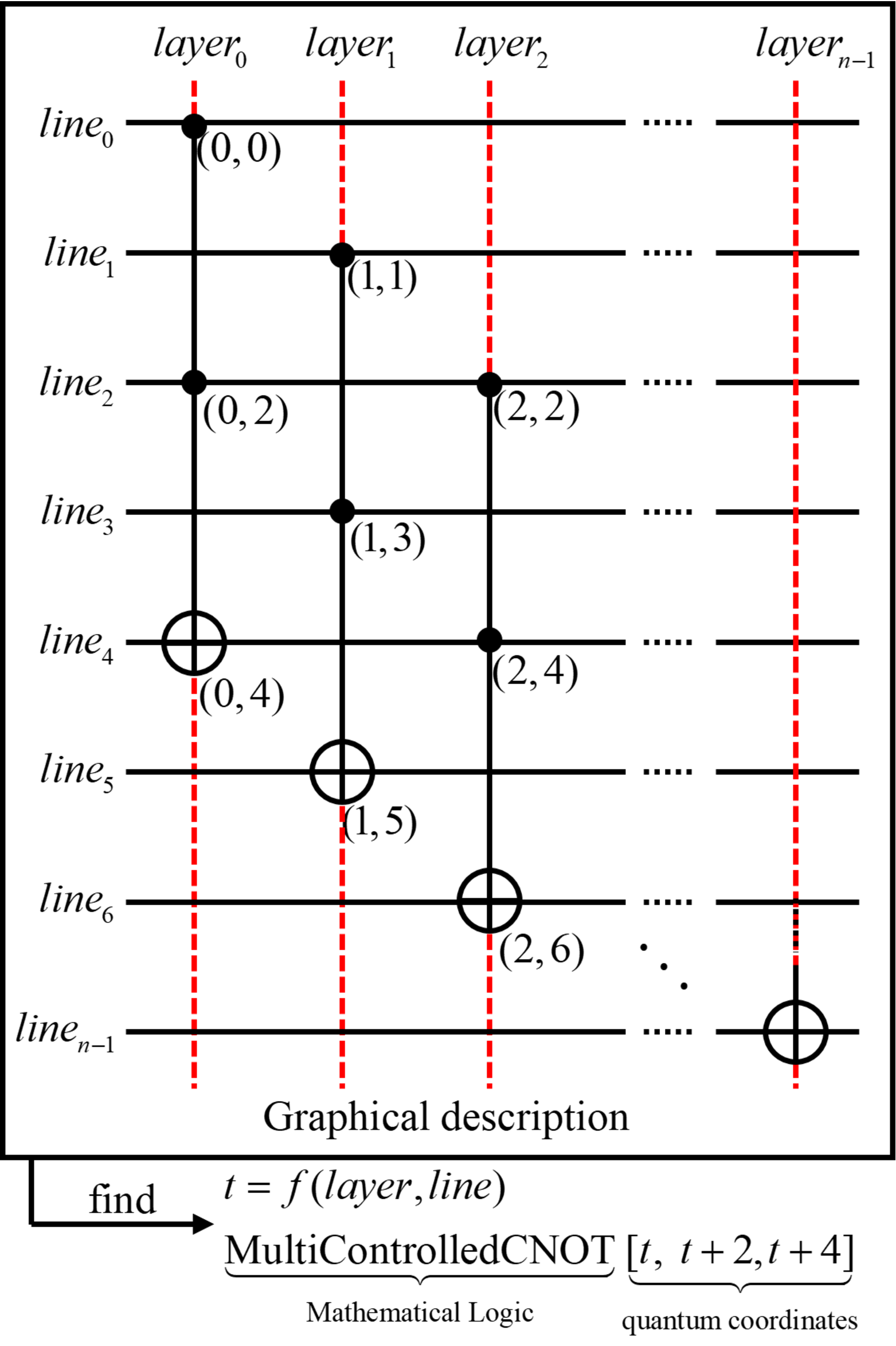}
	\vspace{0em}\caption{An example of a multi-controlled CNOT gate network to illustrate quantum coordinates.}\label{Quantum Coordinates}
\end{figure}

Quantum coordinates are used to dig the mathematical general term between control or output bits to quickly model the network. A simple case is exhibited in Fig. \ref{Quantum Coordinates} to delineate how a quantum circuit diagram with a regular layout can be converted into a mathematical representation.

Intuitively, Fig. \ref{Quantum Coordinates} reveals that there is a special mathematical pattern between the variable line and the variable layer, and between the control and output bits of the multi-controlled CNOT gate. Then the variables \textit{layer} and \textit{line} can be chosen to jointly describe this network, which transforms the graphical language into a mathematical expression.

\begin{equation}\label{ord}
\begin{aligned}
	& t=f(layer, line) \\ 
	& \text{MultiControlledCNOT}[t,t+2,t+4] \\ 
\end{aligned}
\end{equation}
in Fig. \ref{Quantum Coordinates} is the mathematical law of this particular network, where the symbol MultiControlledCNOT represents the logical relationship between the control and output bits, and the contents of the bracket $ [t,t+2,t+4] $ (i.e., quantum coordinates) indicates the specific locations of the control and output bits in this network. Moreover, the mathematical formula in Eq. (\ref{ord}) is obtained by mathematical induction.

Based on the above definition, it is even possible to derive the coordinates of the entire network. Then the whole quantum network can be displayed in the form of coordinate points or can be generalized in a generalized term formula, which enhances the interpretability of the network. The induction by means of coordinate points or generalized terms may provide a feasible solution for quantum circuit optimization. Later on, the charm of quantum coordinates is exhibited. Here, a CNOT gate coordinate law applicable to this project is extracted, which  performs a crucial role. In the same register, the quantum coordinate of the CNOT gate is
\begin{equation}\label{CNOT1}
	CNOT[s(r),s(r)+1],
\end{equation}
where 
\begin{equation}\label{s(t)}
	s(r)=m\times \frac{r-r\,\bmod \,(m-1)}{m-1}+r\,\bmod \,(m-1)
\end{equation}
is a general term formula with respect to $r$. This expression is more concise. The logical function it implies is to XOR the $s(r)$-th and $(s(r) + 1)$-th in the same register. The value range of $r$ depends on the situation.

\subsubsection{Implementation Steps}
9 calculation steps (Step 1-8 and measurements) and the 5 functional modules previously mentioned in the framework are explained here.

\begin{figure}[h]
	\vspace{0em}\centering
	\includegraphics[scale=0.23]{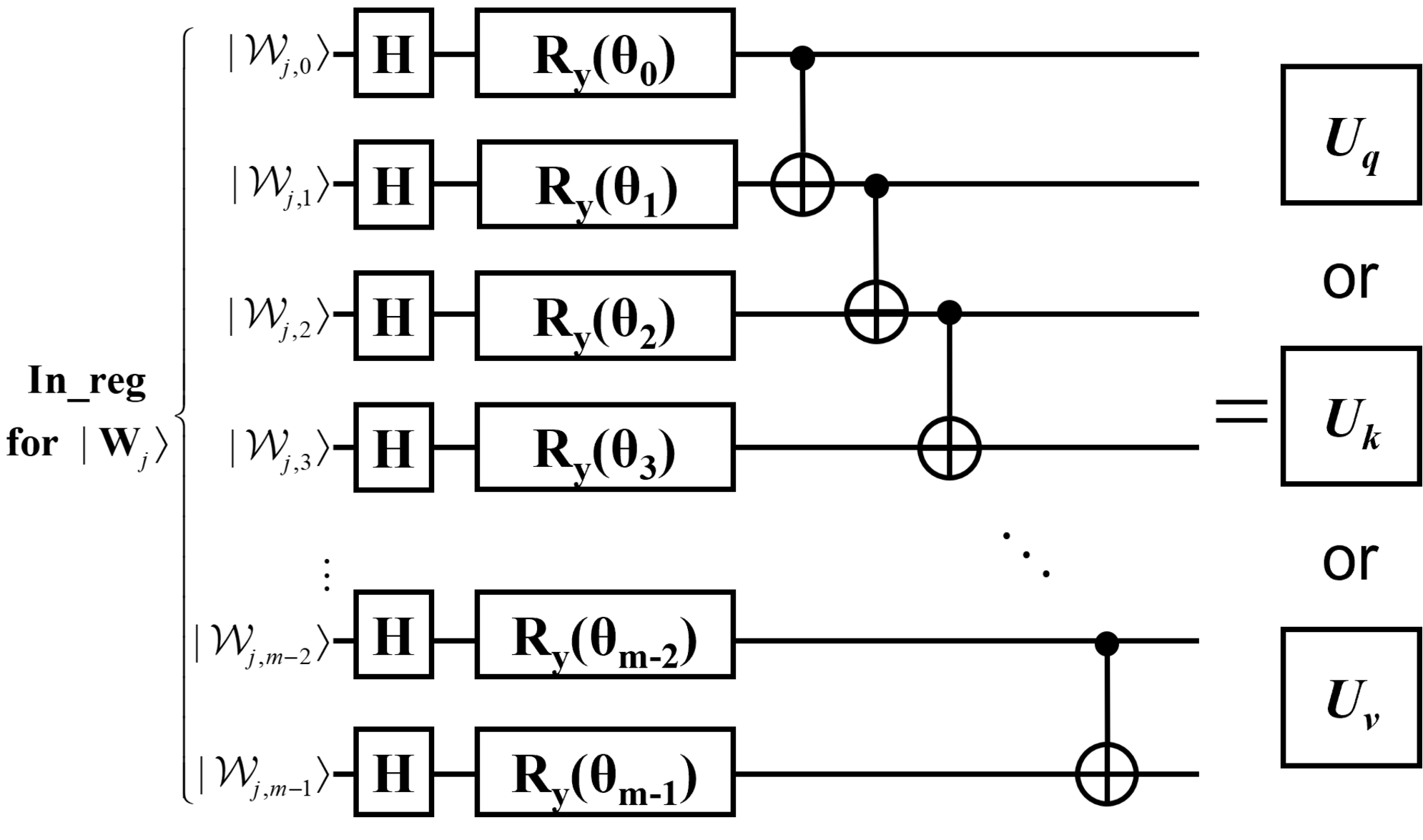}
	\vspace{0em}\caption{Circuit for $U_{q}$ or $U_{k}$ or $U_{v}$}\label{S1}
\end{figure}

\textbf{Step 1}: calculate the query quantum state $|\mathbf{Q}\rangle$ according to Eq. (\ref{Qi}). The procedure is as follows.
\begin{equation}
	|{{U}_{q}}^{\otimes n}\mathbf{In},\mathbf{In},\mathbf{In},\mathbf{0}\rangle =|\mathbf{Q},\mathbf{In},\mathbf{In},\mathbf{0}\rangle,
\end{equation}
where ${{U}_{q}}^{\otimes n}$ is shown in Fig. \ref{S1} in the order provided by Eq. (\ref{UM}).

\begin{figure}[h]
	\vspace{0em}\centering
	\includegraphics[scale=0.19]{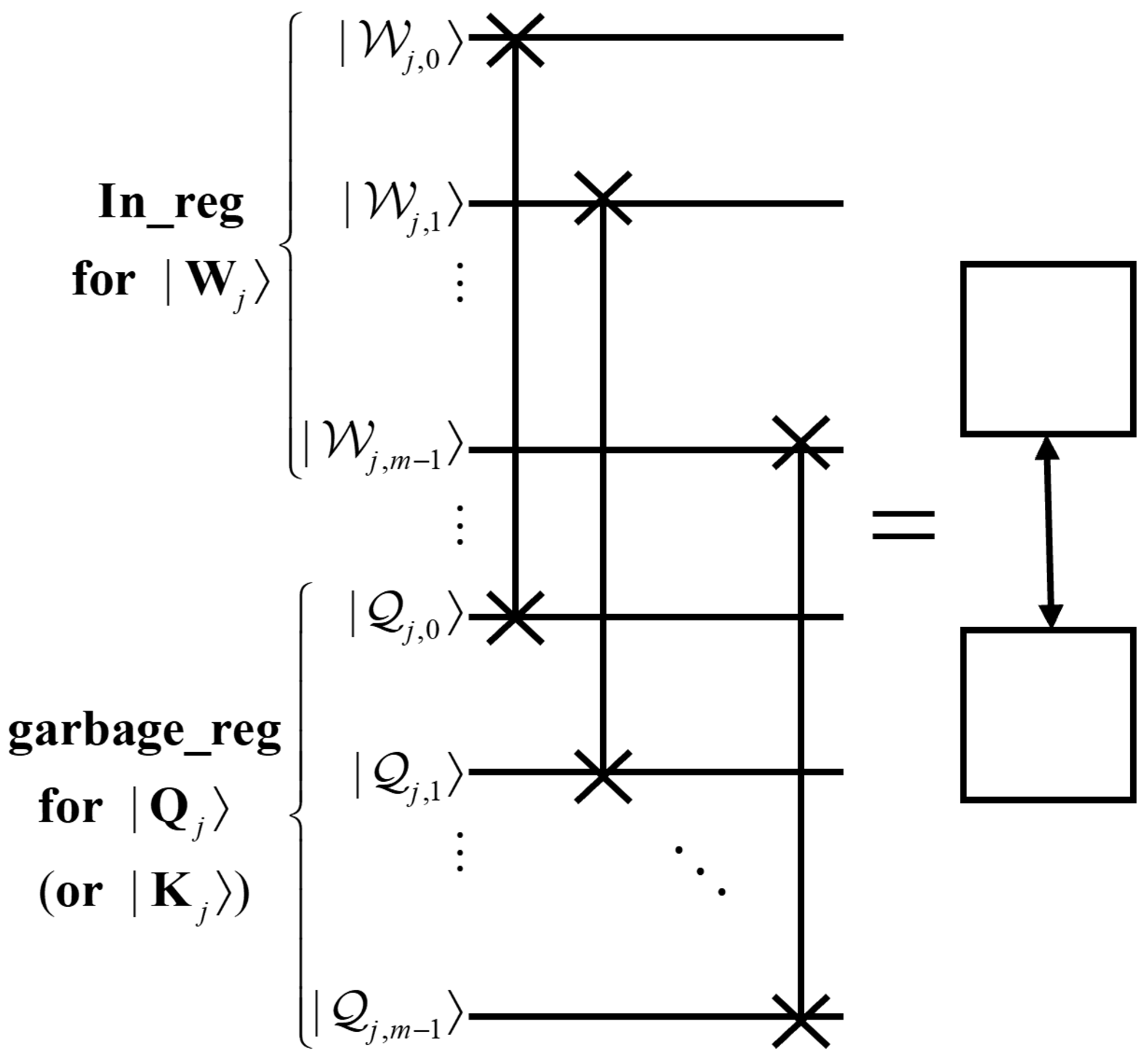}
	\vspace{0em}\caption{Circuit for barbell operation}\label{S2}
\end{figure}
\begin{figure}[h]
	\vspace{0em}\centering
	\includegraphics[scale=0.32]{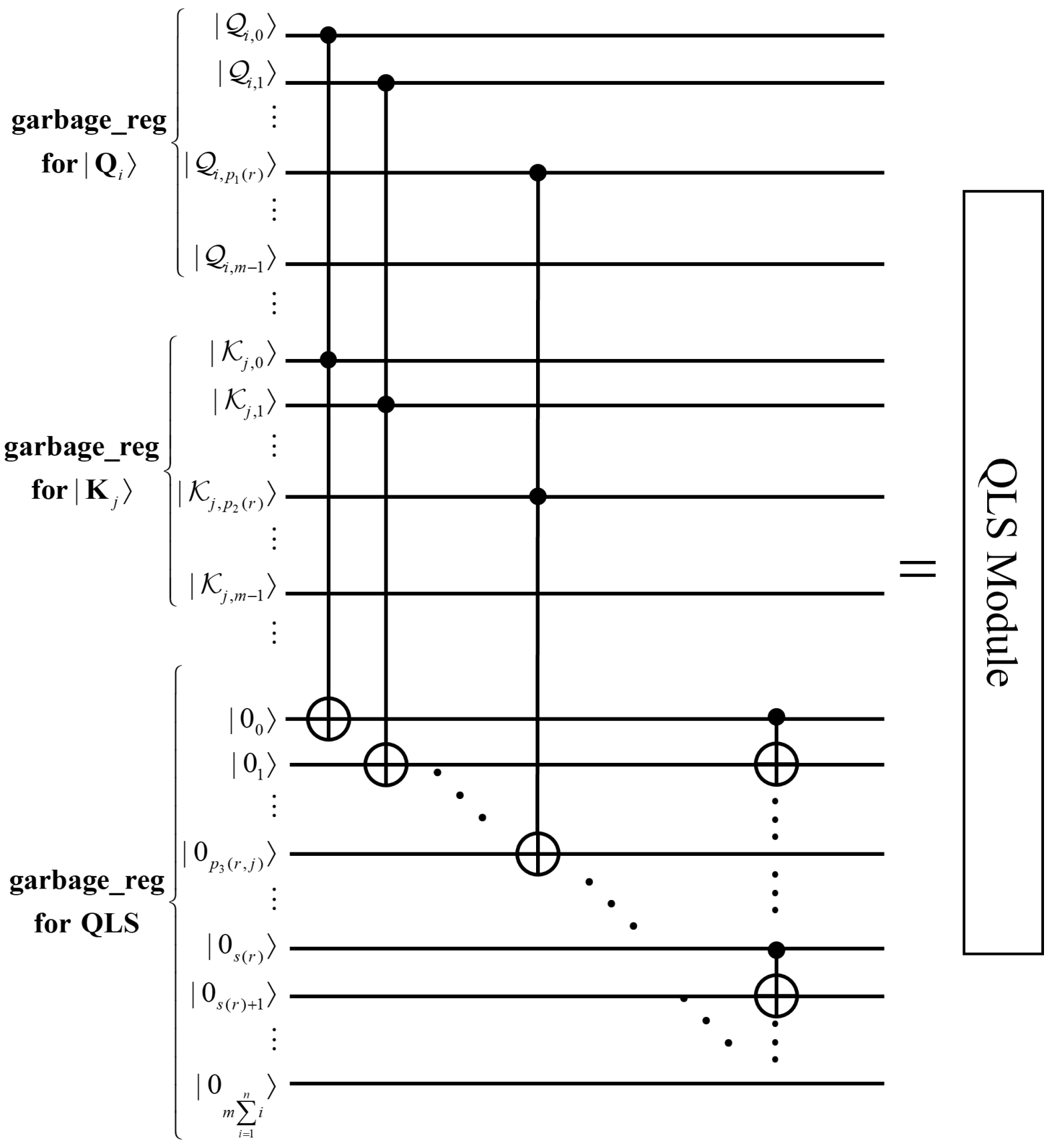}
	\vspace{0em}\caption{Circuit for QLS module}\label{S5}
\end{figure}

\textbf{Step 2}: perform a barbell operation. The barbell operation, which gets its name from the module's form factor, actually swaps the input value of the second garbage register with the current value of the input register. This operation causes the input register to be reset and the result $|\mathbf{Q}\rangle$ to be saved in the second garbage register. The exact procedure is explained by the following equation: 
\begin{equation}\label{step2}
\underset{r}{\mathop{\otimes }}\,SWAP[r,r+m\times n]|\mathbf{Q},\mathbf{In},\mathbf{In},\mathbf{0}\rangle =|\mathbf{In},\mathbf{Q},\mathbf{In},\mathbf{0}\rangle,
\end{equation}
\textcolor{red}{where ${{\otimes }_{r}}SWAP[r,r+m\times n]$ as shown in Fig. \ref{S2} indicates that SWAP gates should be used for each dimension of each output. $SWAP[r,r+m\times n]$ denotes the exchange of the quantum states between the $r$-th and $(r+m\times n)$-th lines. Take Fig. \ref{S2} as an example, i.e. 
	\begin{equation}
		SWAP\text{ }\!\!|\!\!\text{ }{{\mathcal{W}}_{j,i}},{{\mathcal{Q}}_{j,i}}\rangle =\text{ }\!\!|\!\!\text{ }{{\mathcal{Q}}_{j,i}},{{\mathcal{W}}_{j,i}}\rangle,
	\end{equation}
	when $\mathcal{W}_{j,i}$ is at line $i$ and $\mathcal{Q}_{j,i}$ is at line $(i+m\times n)$.}

\textbf{Step 3}: calculate the key quantum state $|\mathbf{K}\rangle$ according to Eq. (\ref{Kj}). The details are shown in Fig. \ref{S1}. The mathematical equation is expressed as
\begin{equation}
	|{{U}_{k}}^{\otimes n}\mathbf{In},\mathbf{Q},\mathbf{In},\mathbf{0}\rangle =|\mathbf{K},\mathbf{Q},\mathbf{In},\mathbf{0}\rangle. 
\end{equation}

\textbf{Step 4}: perform a barbell operation. This time the present data of the input register is exchanged with the content of the third register: 
\begin{equation}
\underset{r}{\mathop{\otimes }}\,SWAP[r,r+2m\times n]|\mathbf{K},\mathbf{Q},\mathbf{In},\mathbf{0}\rangle =|\mathbf{In},\mathbf{Q},\mathbf{K},\mathbf{0}\rangle.  
\end{equation}

\textbf{Step 5}: calculate the QLS according to Eq. (\ref{QK}). The details are drawn in Fig. \ref{S5}.

Firstly, the AND operation is conducted on the qubits in the same position of $|\mathbf{Q}\rangle$ and $|\mathbf{K}\rangle$, and the result is stored in the last garbage register:
\begin{equation}
	Toffol{{i}^{\otimes (m\sum\limits_{i=1}^{n}{i})}}|\mathbf{In},\mathbf{Q},\mathbf{K},\mathbf{0}\rangle =|\mathbf{In},\mathbf{Q},\mathbf{K},\mathbf{Q}\wedge \mathbf{K}\rangle.
\end{equation}

Using the coordinates, $\mathbf{Q}\wedge \mathbf{K}$ is defined as
\begin{equation}\label{Q_AND_K}
	\mathbf{Q}\wedge \mathbf{K}:=\underset{r,j}{\mathop{\otimes }}\,Toffoli[{{p}_{1}}(r),{{p}_{2}}(r),{{p}_{3}}(r,j)],
\end{equation} 
where
\begin{equation}
	{{p}_{1}}(r)=r+m\times n,
\end{equation}
\begin{equation}
	{{p}_{2}}(r)=r+2m\times n,
\end{equation}
\begin{equation}
	{{p}_{3}}(r,j)=r+m\times j+m(\sum\limits_{c=1}^{n-1}{c}-\sum\limits_{d=1}^{n-1-\left\lfloor r/m \right\rfloor }{d})+3m\times n
\end{equation} 
with $r\in \{0,\cdots ,m\times n-1\}$ and $j\in \{0,\cdots ,n-\left\lfloor r/m \right\rfloor -1\}$. $m\times n$, $2m\times n$ and $3m\times n$ are biases for locating which register the current control or output bit is in, e.g. $m\times n$ means it is in the first garbage register and $2m\times n$ represents it is in the second garbage register.

Secondly, the CNOT gates are applied to the fourth garbage register to acquire the eventual result of QLS. According to the law summarized in Eq. (\ref{CNOT1}), the process of applying a CNOT gate at this point is defined as 
\begin{equation}\label{s_r}
	\underset{r}{\mathop{\otimes }}\,CNOT[s(r)+3m\times n,s(r)+3m\times n+1],
\end{equation} 
with $r\in \{0,\cdots ,(m-1)\sum\nolimits_{i=1}^{n}{i}\}$. The fourth register can be located by adding bias $3m\times n$.

The above two steps complete the whole operation steps of QLS: 
\begin{equation}
	|\mathbf{In},\mathbf{Q},\mathbf{K},\langle \mathbf{Q}|\mathbf{K}\rangle \rangle.
\end{equation} 
But the fact is that the outputs $(s(r)+3m\times n+1)$ in Eq. (\ref{s_r}) of QLS do not all need to be concerned. That is, one needs to filter the entire range of values $r\in \{0,\cdots ,(m-1)\sum\nolimits_{i=1}^{n}{i}\}$ from Eq. (\ref{s_r}) to find the part that the project wants. Here, 
\begin{equation}\label{go}
	g(o)=m\times o-1+3m\times n\in (s(r)+3m\times n+1)
\end{equation}  
with $o\in \{1,\cdots ,\sum\nolimits_{i=1}^{n}{i}\}$ is picked from $r\in \{0,\cdots ,(m-1)\sum\nolimits_{i=1}^{n}{i}\}$ as the true QLS output. Once the effective outputs $g(o)$ of QLS are available, the distribution of the outputs can be accessed by programmatically querying the density matrix of $g(o)$, i.e., the by-product QBSASM.

\textbf{Step 6}: calculate the key quantum state $|\mathbf{V}\rangle$ according to Eq. (\ref{Vj}) and Fig. \ref{S1}:
\begin{equation}
	|{{U}_{v}}^{\otimes n}\mathbf{In},\mathbf{Q},\mathbf{K},\langle \mathbf{Q}|\mathbf{K}\rangle \rangle =|\mathbf{V},\mathbf{Q},\mathbf{K},\langle \mathbf{Q}|\mathbf{K}\rangle \rangle. 
\end{equation} 

\begin{figure}[h]
	\vspace{0em}\centering
	\includegraphics[scale=0.22]{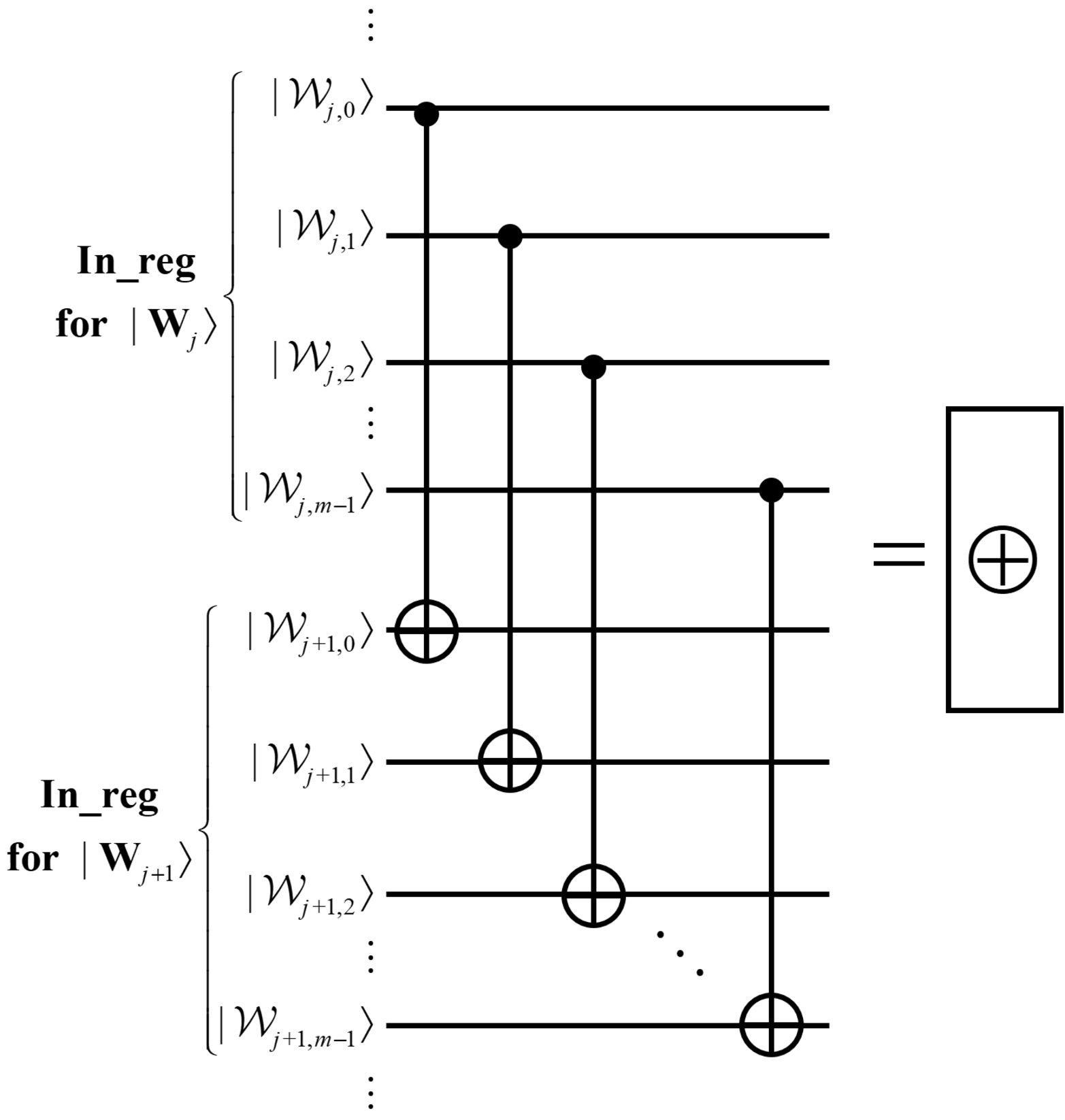}
	\vspace{0em}\caption{Circuit for entanglement compression operation}\label{S6}
\end{figure}

\textbf{Step 7}: perform the entanglement compression operation in Fig. \ref{S6}. This means the output is compressed to the last output of the input register after entanglement by CNOT gates to reduce the number of measurements. 
CNOT gates are added for $|\mathbf{V} \rangle$. The specific way of adding CNOT is executed according to Eq. (\ref{QSA}) and Eq. (\ref{rule}). Specifically, 
\begin{equation}\label{CNOT_V}
	\prod\limits_{m\times n}{{{I}_{{{2}^{i}}}}\otimes CNOT\otimes {{I}_{{{2}^{n-i}}}}}|\mathbf{V}\rangle =\underset{i=0}{\mathop{\overset{m-1}{\mathop{\otimes }}\,}}\,\underset{j=0}{\mathop{\overset{n-1}{\mathop{\oplus }}\,}}\,{{\mathcal{V}}_{i,i+m\times j}}
\end{equation}
if $|{{\mathbf{V}}_{i}}\rangle$ is written as Eq. (\ref{Vjstring}).

\begin{figure}[h]
	\vspace{0em}\centering
	\includegraphics[scale=0.18]{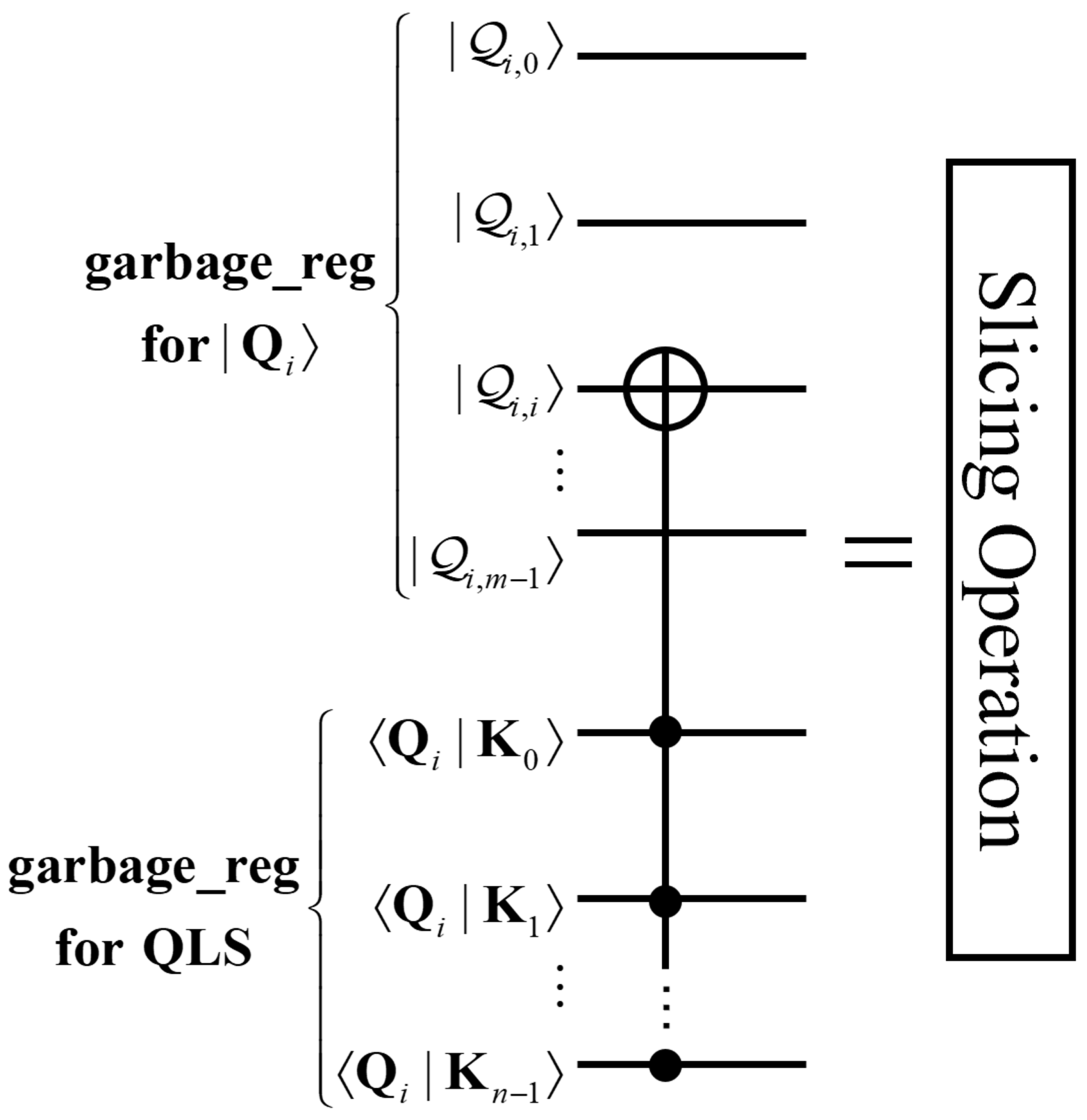}
	\vspace{0em}\caption{Circuit for slicing operation}\label{S8}
\end{figure}

\textbf{Step 8}: execute the slicing operation as shown in Fig. \ref{S8} and select the control bits in accordance with Eq. (\ref{Matrix}). 

First of all, for Fig. \ref{S8}, the Reset operation \cite{29} must be performed before applying the multi-controlled quantum gates, as the original output is not allowed to have any further effect on the present result. Secondly, the relationship between the element $\langle {{\mathbf{Q}}_{{{j}_{1}}}}|{{\mathbf{K}}_{{{j}_{2}}}}\rangle $ of QBSASM and the coordinate $g(o)$ is explored, where ${j}_{1}$ is the row number and ${j}_{2}$ is the column number. Observing Eq. (\ref{Matrix}) and Eq. (\ref{go}), the parameter $o$ of Eq. (\ref{go}) has the following relationship with the positions of the matrix elements:
\begin{equation}
o=\left\{ \begin{array}{*{35}{l}}
	1+{{j}_{2}}+\sum\limits_{i=1}^{n}{i}-\sum\limits_{j=1}^{n-{{j}_{1}}}{j} & {{j}_{1}}\le {{j}_{2}}  \\
	1+{{j}_{1}}+\sum\limits_{i=1}^{n-1}{i}-\sum\limits_{j=1}^{n-1-{{j}_{2}}}{j} & else  \\
\end{array} \right..
\end{equation}
When ${j}_{1}\le {j}_{2}$, ${j}_{1}\in \{0,\cdots ,n-1\}$, ${j}_{2}\in \{0,\cdots ,n-{j}_{1}\}$; otherwise ${j}_{1}\in \{1,\cdots ,n-1\},{j}_{2}\in \{0,\cdots ,{j}_{1}\}$. In this way, the equivalence between the coordinates of the quantum gate and the positions of the elements of the weight matrix is established, then the coordinates of the quantum gate can be confirmed by retrieving the positions of the corresponding elements.

\textbf{Step 9}: combined measurements. This step is measured with skill. Choosing the full output qubit of Eq. (\ref{CNOT_V}) and one of the qubits in Eq. (\ref{go}), the corresponding output can be formed, which also conforms to the reality that the output has 1 more qubit than the input. If the dimensionality is to be guaranteed to be the same, a layer of neural network can be used. 

In general, with the detailed design of the 9 steps, QSAN can not only prevent the measurement in the middle process to ensure the calculation with quantum data until the advent of Step 9, but also compress the number of measurements when Step 9 arrives.
\begin{table*}[]
	\renewcommand{\arraystretch}{1.5}
	\def\tablename{Tab.}
	\centering
	\caption{Method Comparison}
	\label{Method Comparison}
	\resizebox{\textwidth}{!}{%
		
		\begin{tabular}{ccccccc}
			
			\hline
			\multirow{2}{*}{Indicators} & \multicolumn{6}{c}{Models}                                           \\ \cline{2-7} 
			&QSAN & Vaswani’s \cite{0.0}& Niu’s \cite{5.0}& Zhao’s \cite{5.1}& Li’s \cite{5.11} & Qin’s \cite{Q1} \\ \hline
			realizability on quantum computers & completely & $\times$ & $\times$ & $\times$ & partially & $\times$ \\
			
			methods for solving the attention score & QLS  & dot product  &  weak measurement & density matrix  & Gaussian projection  & quantization  \\
			
			data types for attention scores & tensor & scalar  & scalar & scalar & scalar & scalar \\

			
			compressing the number of measurements & $\checkmark$ & - & - & - & $\times$ & -   \\ \hline
		\end{tabular}
	}
\end{table*}
\section{Method Comparison and Experiment}\label{sec4}

In this section, the features of QSAN are highlighted by theoretical comparisons. In addition, the feasibility of QSAN with two cores of this paper (QLS and quantum coordinates) is verified by conducting the following experiments on the IBM Qiskit and pennylane platforms. Specifically, the following analysis and experiments are performed:

\begin{itemize}
	\item Method Comparison: A theoretical comparison between QSAN and Refs. \cite{0.0,5.0,5.1,5.11,Q1}.
	\item Experiment 1: The differences between QLS and common quantum similarity ways (i.e. Hadamard test and Swap test) are compared, highlighting the straightforwardness of QLS to solve for similarity.
	\item Experiment 2: A simple quantum circuit network is constructed, demonstrating that quantum coordinates can convert graphical language to mathematical language, enhancing modeling capabilities and the ability to screen output signals, as well as facilitating programming.
	\item Experiment 3: QSAN is compared with the mainstream hardware-efficient ansatz and QAOA ansatz to demonstrate that QSAN is faster in classifying the MNIST dataset while obtaining the same prediction accuracy.
\end{itemize}

\subsection{Method Comparison}
\textcolor{red}{Presently, freely available quantum computers grapple with severe hardware resource constraints. Furthermore, when QSAN is simulated with a classical system, memory requirements grow exponentially as the number of simulated qubits increases. 
Thus QSAN is theoretically compared with Refs. \cite{0.0,5.0,5.1,5.11,Q1} to highlight its characteristics in terms of four aspects: implementability on quantum computers, methods for realizing attention scores, data properties of attention scores, and measurement compression. The specific comparison results are shown in Tab. \ref{Method Comparison}.}

\textcolor{red}{From a QML perspective, Tab. \ref{Method Comparison} lists several important indicators. Firstly, both QSAN and Li's method \cite{5.11} are implementable on current quantum computing platforms, underscoring the positive effect of QSAN in the QML area. The difference is that QSAN can be completely on a quantum computer, i.e., the self-attention score and the output can be derived from its quantum circuit simultaneously. In contrast, there are additional classical processing steps involved in Li's method. Refs. \cite{5.0, 5.1} utilize quantum physics concepts to improve classical model performance without designing specific quantum circuits. Refs. \cite{0.0,Q1} belong to the classical machine learning domain. QSAN assimilates numerous concepts from Ref. \cite{0.0}, including Key-Query-Value and self-attention score, whlie leverages these concepts to derive a novel operational mechanism without strictly adhering to the framework of Ref. \cite{0.0}.
Meanwhile, Ref. \cite{Q1} employs quantization to logicize a portion of the operation of SAM, but distinctions persist between digital and quantum logic. For example, quantum logic, founded on tensor operations in Hilbert space, provides a more precise characterization of the physical properties of particles. Additionally, rotating quantum gates with parameters are more flexible than digital logic, boosting applicability.}

\textcolor{red}{Another discrepancy arises in the approach to solving the self-attention scores and theutilized data type. Despite the diverse array of methods for solving self-attention scores, Refs. \cite{0.0,5.0,5.1,5.11,Q1} conceptualize the self-attention score as a scalar, while QSAN consistently upholds it as a tensor in the Hilbert space. In other words, QSAN always maintains the self-attention score as a quantum state by virtue of QLS. This endows QSAN with probabilistic self-attention scores, significantly expands the data representation space of QBSASM beyond the classical self-attention score matrix, and establishes the groundwork for QSAN to obtain outputs and self-attention scores concurrently in a single step. }

\textcolor{red}{Finally, QSAN takes into account the compression for the number of measurements. Drawing from the VQA framework in subsection \ref{vqa}, outcomes of QSAN are ultimately extracted via quantum measurements. The combined measurements at step 9 in Fig. \ref{QSAN} can compress the storage of the results of quantum measurements in a classical computer. Reducing the number of measurements can somehow minimize the effect of quantum channel noise and save classical storage at the same time.}

\subsection{Experiment 1: Comparison of Quantum Logical Similarity and Quantum State Overlap Circuits}

Inspired by CRC checksums and Frobenius inner products, QLS is an uninterrupted unsigned sum that aims to compress information into a qubit.

\begin{figure}[]
	\vspace{0em}\centering
	\includegraphics[scale=0.22]{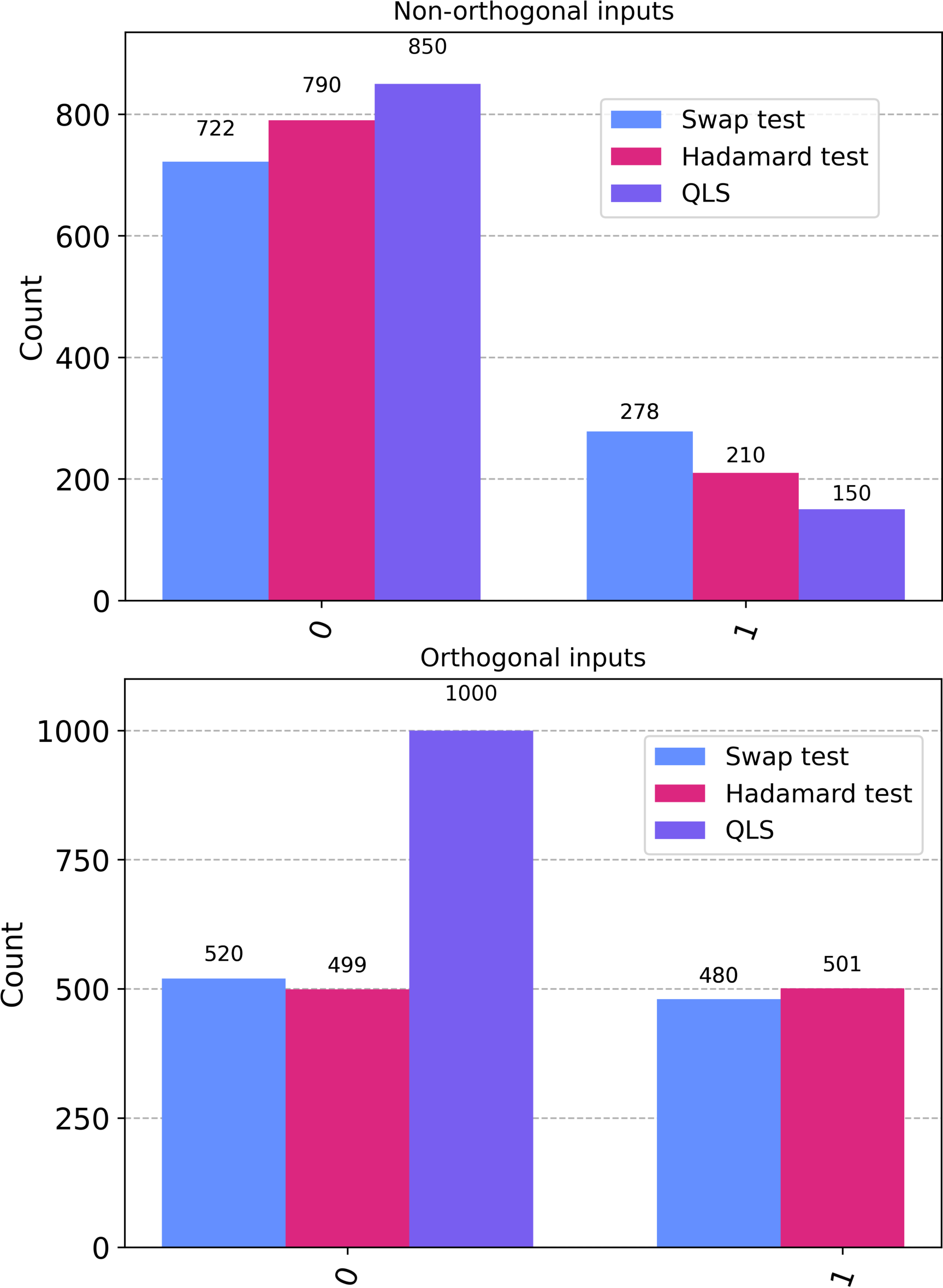}
	\vspace{0em}\caption{Comparison of results for QLS and quantum state overlapping circuits.}\label{Comparehada}
\end{figure}

\textcolor{red}{Suppose a pair of non-orthogonal quantum states
\[\begin{aligned}
	 |A\rangle =[&0.8653396721770911,0.12078742375711317, \\ 
	& 0.10569498331118206,0.4747907123369339{{]}^{\text{T}}} \\ 
\end{aligned}\]
and 
\[\begin{aligned}
	 |B\rangle =[&0.5322823910118109,0.6874987233326855, \\ 
	& 0.4857139917423643,0.0900159978024791{{]}^{\text{T}}}, \\ 
\end{aligned}\]
and a pair of orthogonal quantum states
$$|C\rangle ={{[1,0,0,0]}^{\text{T}}}$$ and $$|D\rangle ={{[0,1,0,0]}^{\text{T}}}$$ are randomly generated. It is also known that the mathematical relation between similarity and measured probability is demonstrated as 
\begin{equation}\label{HadaT}
	P(|0\rangle )=\frac{1+\operatorname{Re}(\langle stat{{e}_{1}}|stat{{e}_{2}}\rangle )}{2}
\end{equation}
on the Hadamard test circuit and 
\begin{equation}\label{SwapT}
	P(|0\rangle )=\frac{1+|\langle stat{{e}_{1}}|stat{{e}_{2}}\rangle {{|}^{2}}}{2}
\end{equation} on the Swap test, where $P(|0\rangle )$ is the measured probability on the ground state $|0\rangle$. $\operatorname{Re}(\langle stat{{e}_{1}}|stat{{e}_{2}}\rangle )$ and $|\langle stat{{e}_{1}}|stat{{e}_{2}}\rangle {{|}^{2}}$ are the similarities, especially $\operatorname{Re}(\langle stat{{e}_{1}}|stat{{e}_{2}}\rangle )$ is also the signed similarity.
Then these two pairs of quantum states are taken as inputs to Hadamard test circuit, Swap test circuit and QLS module, respectively. The outcomes of comparing these three quantum circuits in solving the quantum state similarity are depicted in Fig. \ref{Comparehada}. The horizontal coordinates 0 and 1 indicate ground state $|0\rangle$ and excited state $|1\rangle$. The vertical coordinates indicate the number of statistics, which can be also equivalently viewed as the corresponding probability generated by measuring 1000 times. According to Fig. \ref{Comparehada}, the measurement probabilities of Hadamard test, Swap test and QLS for $|0\rangle$ are 0.722, 0.79 and 0.85, respectively, when the inputs are $|A\rangle$ and $|B\rangle$. Conversely, the probabilities are 0.52, 0.499 and 1 for inputs $|C\rangle$ and $|D\rangle$, respectively.}

\textcolor{red}{Firstly, comparing Eq. (\ref{QK}), Eq. (\ref{HadaT}), and Eq. (\ref{SwapT}), QLS is more straightforward relative to the Hadamard test and the Swap test. The reason is that the output probability of the QLS directly corresponds to the similarity, while for the Hadamard test and the Swap test, the similarity should be derived with an inverse solving process. Secondly, QLS is designed to preserve quantum states without the requirement for measurement during QSAN computation. Consequently, in this experiment, QLS behaves as $\sqrt{0.85}|0\rangle +\sqrt{0.15}|1\rangle $ for non-orthogonal quantum states and $|0\rangle$ for orthogonal quantum states. Mathematically, QLS is a tensor applicable to quantum linear systems. On the contrary, the similarity of Hadamard (or Swap) test is a scalar, since it has to be gained by collapsing the quantum state in measurement. During processing, its similarity is often determined by $\max [P(|0\rangle ),P(|1\rangle )]$.}

\textcolor{red}{In conclusion, compared to the other two similarity methods, QLS is a tensor in mathematical expression and more straightforward in principle, which lays the groundwork for the one-step generation for output and QBSASM of QSAN.}

\subsection{Experiment 2: Modeling and Screening Capabilities of Quantum Coordinates}

\begin{figure}[]
	\vspace{0em}\centering
	\includegraphics[scale=0.28]{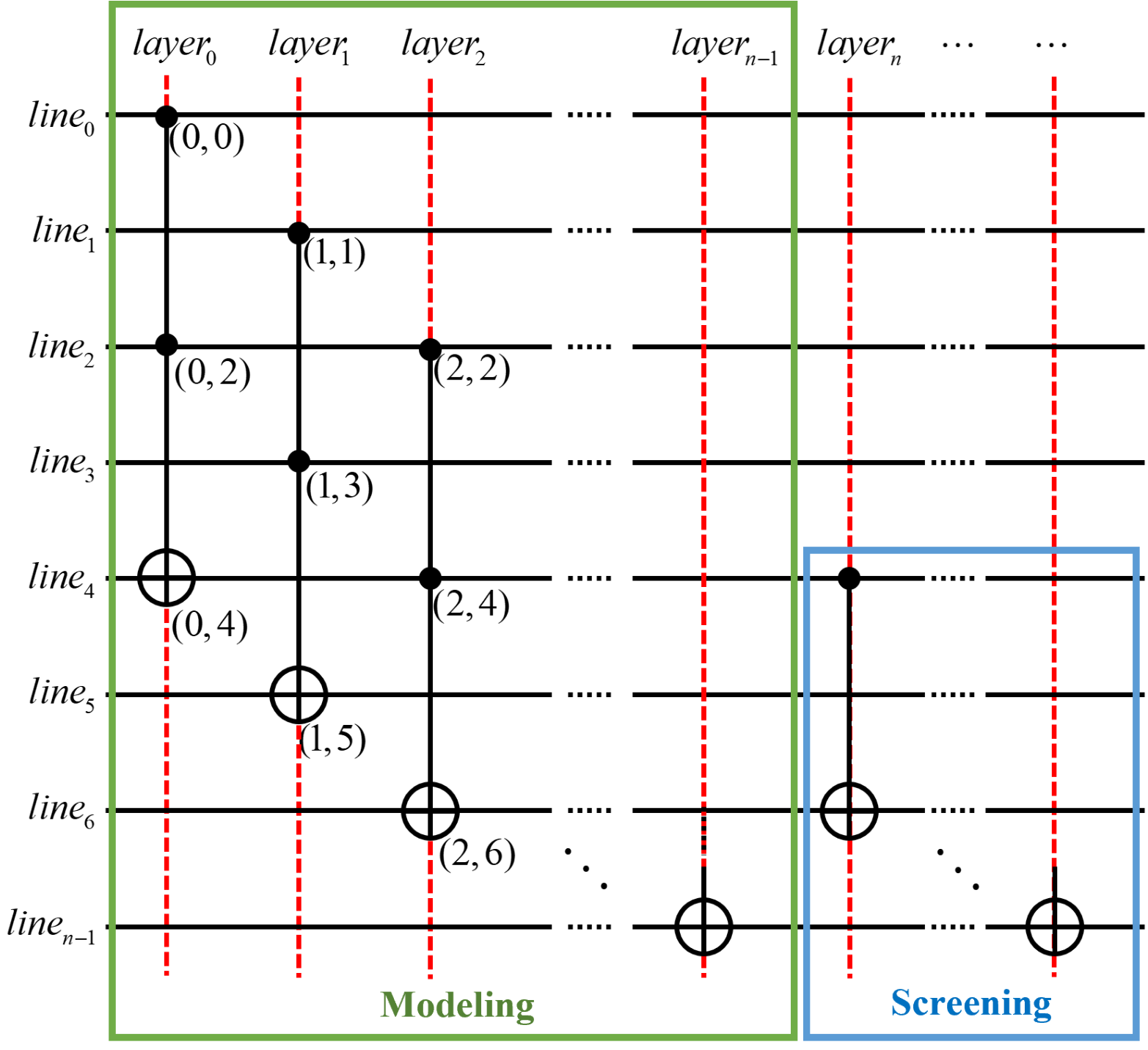}
	\vspace{0em}\caption{Extension of Fig. \ref{Quantum Coordinates} to explain the modeling and screening capabilities of quantum coordinates.}\label{QuantumCoordinates1}
\end{figure}

Quantum coordinates, as the design guideline for QSAN, bring convenience to modeling and filtering output. Here, the structure of Fig. \ref{Quantum Coordinates} is extended, as shown in Fig. \ref{QuantumCoordinates1}, to illustrate this advantage.
\begin{figure}[]
	\vspace{0em}\centering
	\includegraphics[scale=0.2]{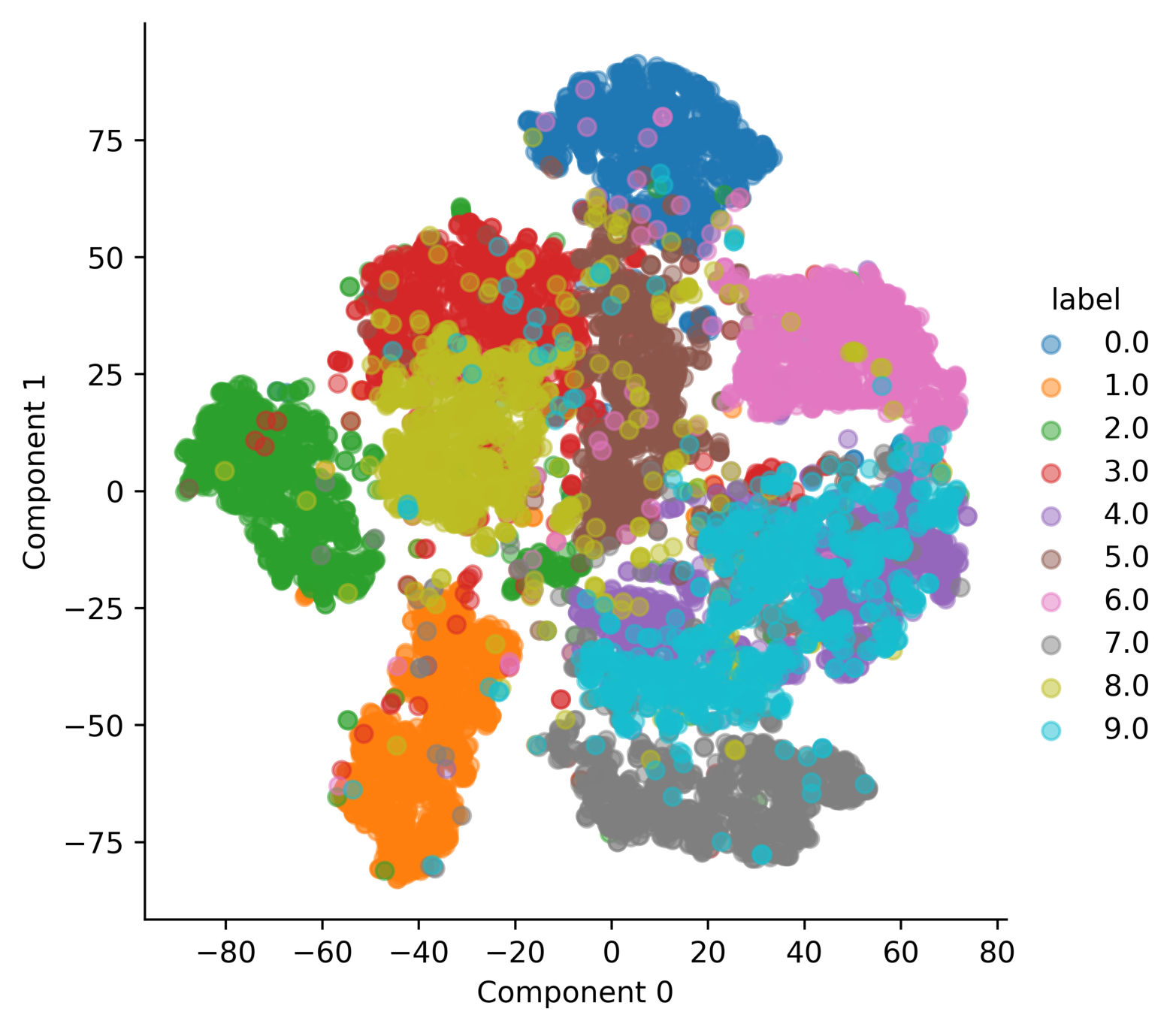}
	\vspace{0em}\caption{Visualization of MNIST dataset}\label{origin}
\end{figure}
\begin{figure*}[]
	\vspace{0em}\centering
	\includegraphics[scale=0.2]{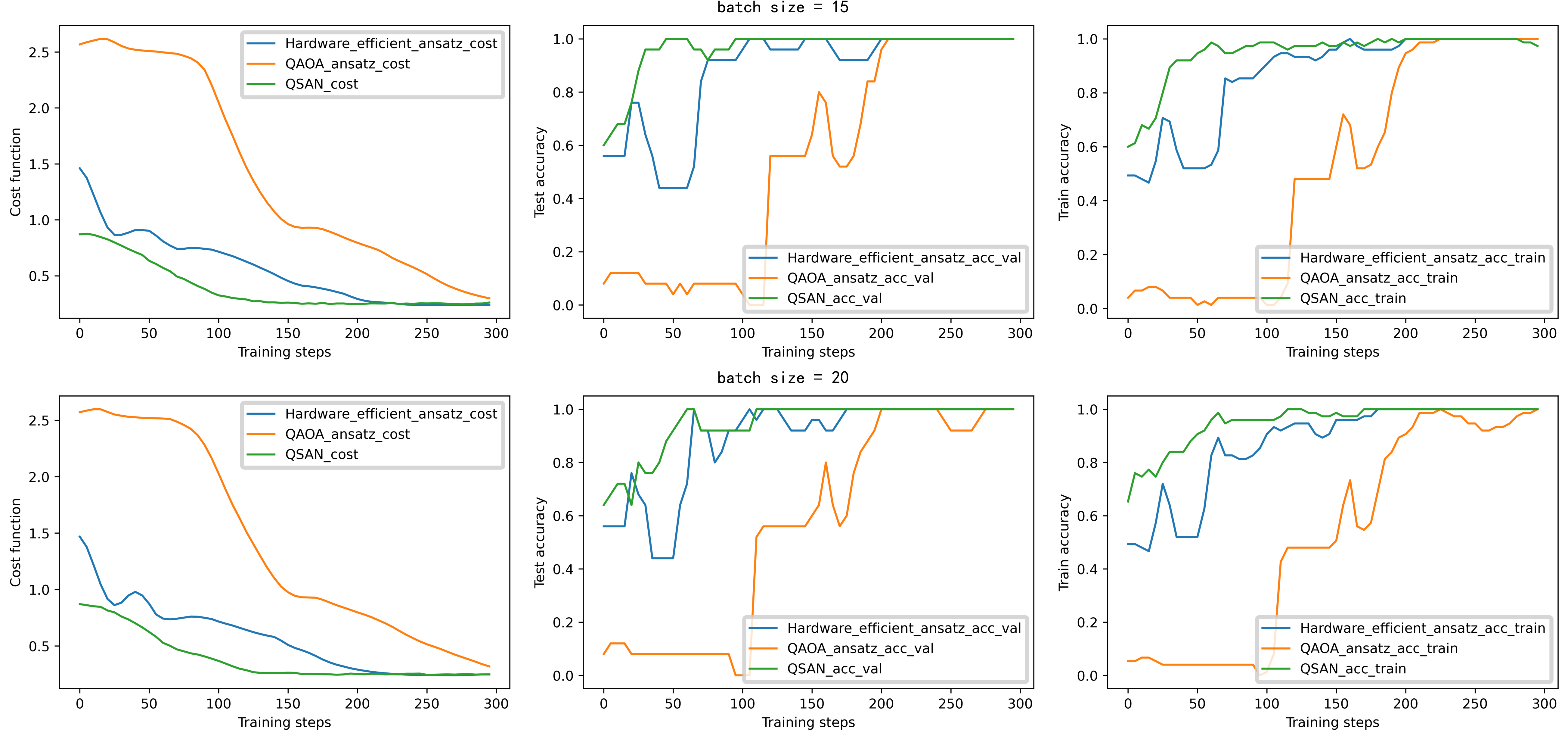}
	\vspace{0em}\caption{Classification results of MNIST data by QSAN ansatz, hardware-efficient ansatz and QAOA ansatz.}\label{compareAnsatz}
\end{figure*}
\begin{figure}[]
	\vspace{0em}\centering
	\includegraphics[scale=0.25]{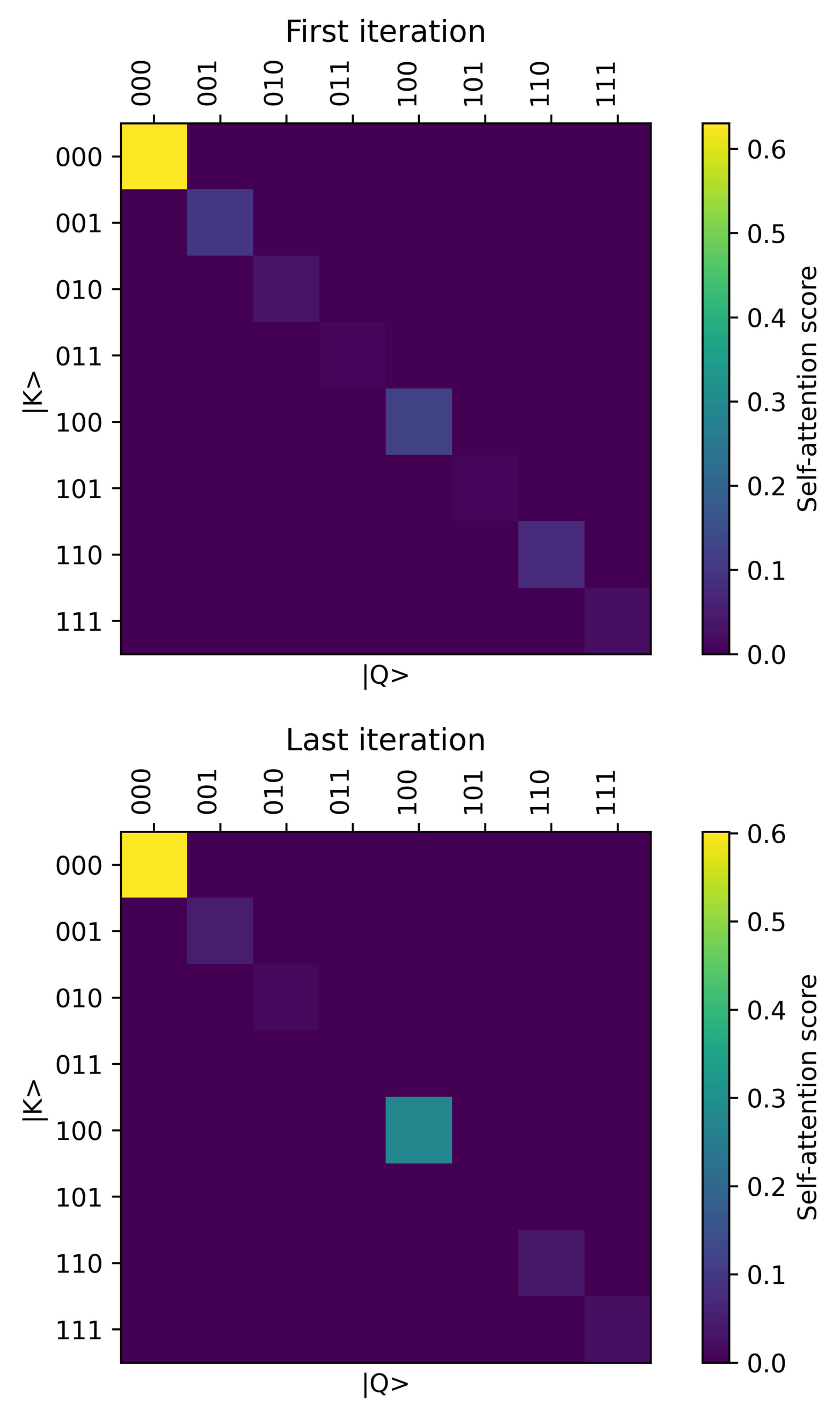}
	\vspace{0em}\caption{Quantum attention score matrix}\label{attention matrix}
\end{figure}

As shown in Fig. \ref{QuantumCoordinates1}, the Toffoli gates form a regular arrangement. The control and output bits of the Toffoli gates located at different layers both intersect the line to produce Cartesian coordinate points, which is the original idea of quantum coordinates.
 
\textit{\textbf{Modeling}}: Following the Cartesian coordinates of Fig. \ref{QuantumCoordinates1}, we can know that the control and output bits of the Toffoli gates are an arithmetic progression. Therefore, the formula 
\begin{equation}\label{AA}
\underset{layer}{\mathop{\otimes }}\,Toffoli[layer,layer+2,layer+4]
\end{equation}
can be obtained by mathematical induction to summarize the laws of the \textit{Modeling} part of Fig. \ref{QuantumCoordinates1}. In Eq. (\ref{AA}), $layer$ (or $layer+2$) denotes the position of the first (or second) control bit. $layer+4$ is the position of the output bit, which is also the position we are interested in. 

\textit{\textbf{Screening}}: Once the coordinates of the output bits are known, the results can be screened. Since in many cases some of the information in the output would be processed again, this makes it necessary to pick out the signals we care about. For example, in this experiment, the output bits with odd serial numbers are picked and then a CNOT gate is applied between them.

According to Eq. (\ref{AA}), the sequence 
\begin{equation}\label{AB}
	\{layer+4\}_{layer=0}^{n-1}
\end{equation}
containing odd and even numbers is obtained. Here the selection operator $g(o)$ is imposed so that the sequence (\ref{AB}) contains only odd numbers. By observation, when $g(o)$ takes an odd number, it makes the condition hold. That is 
\begin{equation}\label{AC}
	g(o)=2o+1,o\in \mathbb{N},g(o)\in layer.
\end{equation}
When the filtering operator is applied, the layers of the function are deepened and the independent variable becomes the $o$ of the filtering operator instead of the $layer$, because $g(o)$ is a layer value in the sense. In the end, the screened model 
\begin{equation}\label{AD}
\underset{o}{\mathop{\otimes }}\,CNOT[g(o),g(o+1)]
\end{equation}
is derived. 

Thus in summary, the final mathematical model of this network structure is as follows.
\begin{equation}\label{AE}
	\begin{aligned}
		& Modeling:\underset{i}{\mathop{\otimes }}\,Toffoli[layer,layer+2,layer+4] \\ 
		& Screening:\underset{o}{\mathop{\otimes }}\,CNOT[g(o),g(o+1)]. \\ 
	\end{aligned}
\end{equation}

This example confirms that quantum coordinates can facilitate programming by converting graphical structures into a mathematical language and providing an interface for modeling the next layer by screening operators.

\subsection{Experiment 3: Classification of MNIST dataset based on QSAN}

There is evidence in the classical domain that SAM can be used for image classification \cite{0.32,0.313}. In this experiment, QSAN, hardware-efficient ansatz and QAOA ansatz are used on the pennylane platform to simultaneously biclassify images 0 and 1 from the MNIST dataset to demonstrate the power of quantum classifiers.

\subsubsection{Pre-processing of the MNIST dataset}
MNIST is a well-known database of handwritten numbers covering 10,000 test images and 60,000 training images \cite{30.1}. Each image has $28\times28$ pixel points. However, due to the limitation of the number of qubits in the NISQ era, this dataset must be preprocessed. Preprocessing here refers to some dimensionality reduction operations. 

After the size of the original image to $4\times4$, these data are visualized in Fig. \ref{origin}. The labels in Fig. \ref{origin} indicate the numeric type of handwriting. In this experiment, only the data with labels 0 and 1 are applied. Further, the first 50 (or 30) data of label 0 and 1 are selected as the training (or test) set, respectively.

\subsubsection{Experimental settings}
In order to verify the effectiveness of QSAN, two typical models in QML, hardware-efficient ansatz \cite{5.9} and QAOA ansatz \cite{5.12} are chosen for comparison. 
They process the same MNIST dataset. The specific experimental parameter settings are shown in Tab. \ref{Experimental comparison}, where the training parameters, such as learning rate, optimizer, etc. are kept consistent. \begin{table}[]
	\renewcommand{\arraystretch}{1.5}
	\def\tablename{Tab.}
	\centering
	\caption{Experimental configuration}
	\label{Experimental comparison}
	\resizebox{\columnwidth}{!}{%
		\begin{tabular}{cccc}
			\hline
			\multirow{2}{*}{Indicators} & \multicolumn{3}{c}{Models}                                           \\ \cline{2-4} 
			&Hardware-efficient \cite{5.9}&QAOA \cite{5.12}&QSAN \\ \hline
			
			
			parameters    & 12                        & 16                    & 12                        \\
			
			layers        & 3                         & 1                      & 1                         \\
			
			data encoding & amplitude                 & angle                  & amplitude                 \\
			
			representation advantages & \checkmark                 & $\times$                 & \checkmark                 \\
			
			attention score & $\times$                & $\times$                 & \checkmark                 \\
			
			learning rate & \multicolumn{3}{c}{0.02}                                                       \\
			
			batch\_size   & \multicolumn{3}{c}{15; 20}                                                          \\
			
			step          & \multicolumn{3}{c}{300}                                                        \\
			
			optimizer     & \multicolumn{3}{c}{NesterovMomentumOptimizer}                                  \\
			\hline
		\end{tabular}%
	}
\end{table} However, the parameters, number of layers and data encoding methods vary due to the different quantum models.

\subsubsection{Experimental analysis}

\textbf{\textit{Training results}} The classification results of the three models for MNIST data are shown in Fig. \ref{compareAnsatz}. 
The blue line, orange line, and green line indicate hardware-efficient ansatz, QAOA ansatz, and QSAN, respectively. The first, second and third plots in Fig. \ref{compareAnsatz} represent the variation of loss function, test accuracy and training accuracy, respectively. Combining all the plots of experimental results in Fig. \ref{compareAnsatz}, the following conclusions are drawn:

\begin{itemize} 
	\item Quantum circuit model for MNIST dataset classification has 100\% test and training accuracy.
	\item With the same experimental configuration and similar number of model parameters, QSAN begins to converge at step 130, relative to step 220 for hardware-efficient ansatz and step 300 for QAOA ansatz. QSAN converges about 1.7x and 2.3x faster than hardware-efficient ansatz and QAOA ansatz. Additionally, QSAN achieves a lower loss function value, indicating stronger learning capability.
	\item \textcolor{red}{The convergence speed of QSAN is relatively sensitive to the parameter batch size within a certain range of variations. As can be seen from the experimental results in Fig. \ref{compareAnsatz}, when the batch size changes from 15 to 20, the convergence of the cost function, training accuracy and test accuracy curves of QSAN becomes significantly slower.}
\end{itemize}

\textbf{\textit{QBSASM}} These heat maps in Fig. \ref{attention matrix} are the density matrix QBSASM intercepted by pennylane. 
Their axes represent the quantum states $|{{\mathbf{Q}}_{i}}\rangle $, $|{{\mathbf{K}}_{j}}\rangle $ and attention score, respectively. By intercepting the results of the first iteration and the final iteration, it is found that the QBSASM of the output has changed. 

\begin{itemize} 
		\item The high-dimensional property of the QBSASM in Eq. (\ref{Matrix}) under Hilbert space and the exponential information characterization ability are exhibited.
		\item As can be expected, this matrix is difficult to simulate for classical computers because it grows exponentially with increasing input.
\end{itemize}

In conclusion, the above two points demonstrate that QSAN has faster training speed and higher dimensionality of QBSASM. As the dimensionality of outputs rises, QBSASM is difficult to be simulated by classical computers, reflecting the advantage of state space representation of quantum computers.

\section{Conclusion}\label{sec5}

A novel QSAM framework and its practical model QSAN are proposed. QSAM consists of two major parts, QLS and QBSASM. QLS replaces the practice of inner product similarity and avoids the construction of large quantum numerical operation networks, thus saving more qubits and making QSAN fully deployable on quantum computers in one step. QBSASM can be obtained during the evolution of QSAN to present quantum attention distribution in the form of density matrix. QSAN is a practical model of QSAM, divided into 9 specific steps and 5 special functional modules, which allow to obtain QBSASM during the evolution, as well as to compress the number of measurements.
It is worth mentioning that QSAN belongs to a network structure with regular layout. Therefore, quantum coordinates are able to obtain mathematical connections between quantum gate control bits and output bits by induction, which enables to describe the network with mathematical formulas, facilitating the programming implementation and possibly laying the foundation for optimization. In the end, MNIST data binary classification experiments demonstrate that QSAN converges about 1.7x and 2.3x faster than hardware-efficient ansatz and QAOA ansatz with similar configurations, which predicts that the QSAN model has better learning capability. In addition, QSAN, as an extensible module, can be embedded into classical or QML architectures, facilitating the construction of a quantum version of Transformer and laying the foundation for quantum-enhanced machine learning.

\setcounter{table}{0}   


\renewcommand{\thetable}{A\arabic{table}}
\ifCLASSOPTIONcompsoc



\ifCLASSOPTIONcaptionsoff
  \newpage
\fi

\end{document}